\documentclass[aps,prb,reprint,nofootinbib,twocolumn,superscriptaddress,showpacs,showkeys,longbibliography,amsmath,amssymb]{revtex4-2}
\usepackage{graphicx}
\usepackage{dcolumn}
\usepackage{bm}
\usepackage{braket}
\usepackage{subfigure}
\usepackage{color}
\usepackage[colorlinks,bookmarks=false,citecolor=blue,linkcolor=red,urlcolor=blue]{hyperref}
\usepackage[english]{babel}
\usepackage{amssymb}
\usepackage{epstopdf,epsfig}
\usepackage{lipsum}
\usepackage{times} 
\usepackage{helvet} 
\usepackage{courier} 

\makeatletter

\newcommand{\Rmnum}[1]{\expandafter\@slowromancap\romannumeral #1@}
\makeatother

\begin{document}
\preprint{APS/123-QED}

\title{First-order Quantum Phase Transitions and Localization in the 2D Haldane Model with Non-Hermitian Quasicrystal Boundaries} 

\author{Xianqi Tong}
\altaffiliation{Department of Physics, Beijing Normal University, Beijing 100000, People’s Republic of China}
\author{Su-Peng Kou}%
\email{spkou@bnu.edu.cn}
\affiliation{Department of Physics, Beijing Normal University, Beijing 100000, People’s Republic of China}

\date{\today}

\begin{abstract}
The non-Hermitian extension of quasicrystals (QC) are highly tunable system for exploring novel material phases. While extended-localized phase transitions have been observed in one dimension, quantum phase transition in higher dimensions and various system sizes remain unexplored. Here, we show the discovery of a new critical phase and imaginary zeros induced first-order quantum phase transition within the two-dimensional (2D) Haldane model with a quasicrystal potential on the upper boundary. 
Initially, we illustrate a phase diagram that evolves with the amplitude and phase of the quasiperiodic potential, which is divided into three distinct phases by two critical boundaries: phase (I) with extended wave functions, PT-restore phase (II) with localized wave functions, and a critical phase (III) with multifunctional wave functions. To describe the wavefunctions in these distinct phases, we introduce a low-energy approximation theory and an effective two-chain model. 
Additionally, we uncover a first-order structural phase transition induced (FOSPT) by imaginary zeros. As we increase the size of the potential boundary, we observe the critical phase splitting into regions in proportion to the growing number of potential zeros. Importantly, these observations are consistent with groundstate fidelity and energy gap calculations. Our research enhances the comprehension of phase diagrams associated with high-dimensional quasicrystal potentials, offering valuable contributions to the exploration of unique phases and quantum phase transition.
\end{abstract}

\maketitle

\section{Introductioon}
The exploration of open systems, characterized by non-Hermitian quantum systems, has unraveled intriguing phenomena absent in their Hermitian counterparts \cite{non-Hermitian1,non-Hermitian2, non-Hermitian3, non-Hermitian4, non-Hermitian5, non-Hermitian6, non-Hermitian7, non-Hermitian8, non-Hermitian9, non-Hermitian10, non-Hermitian11, non-Bloch_BBC1, non-Hermitian12}. Notable examples include PT symmetry and exceptional points \cite{non-Hermitian4, non-Hermitian5, non-Hermitian6, non-Hermitian7, non-Hermitian8, non-Hermitian9, non-Hermitian10, non-Hermitian11}, non-Bloch bulk-boundary correspondence \cite{non-Hermitian11, non-Bloch_BBC1}, and non-Hermitian skin effects \cite{non-Hermitian11, non-Bloch_BBC1, non-Hermitian_skin_effects1, non-Hermitian_skin_effects2, non-Hermitian_skin_effects3, non-Hermitian_skin_effects4, non-Hermitian_skin_effects5, non-Hermitian_skin_effects6, non-Hermitian_skin_effects7, non-Hermitian_skin_effects8, non-Hermitian_skin_effects9}. Many of these phenomena are related to parity-time (PT) symmetric Hamiltonians. These Hamiltonians typically exhibit two phases as parameters vary: the PT-symmetric phase with real eigenvalues and the PT-breaking phase with complex eigenvalues \cite{non-Hermitian2, non-Hermitian3, non-Hermitian4, non-Hermitian5}. These phenomena have been experimentally observed in open systems \cite{experiment1, experiment2, experiment3, experiment4, experiment5, experiment6, experiment7, experiment8}, with promising applications in precision measurements, nonreciprocal quantum devices, and topological transport. The higher-order non-trivial interplay between the non-Hermitian skin effect and the topological effect has led to the concept of a hybrid skin-topological effect \cite{hybrid_skin-topological_effect4, hybrid_skin-topological_effect5, hybrid_skin-topological_effect6, hybrid_skin-topological_effect1, hybrid_skin-topological_effect2, hybrid_skin-topological_effect3}. 

Quasicrystals (QC) in closed quantum systems exhibit a plethora of fascinating properties \cite{AAH2, AAH3, AAH4, AAH5, AAH6, AAH7, AAH8, AAH9, AAH10}. For instance, in the one-dimensional (1D) Aubry-André-Harper (AAH) model, the introduction of finite quasiperiodic strength leads to a transition from a metallic (extended)  state to an Anderson insulator (localized) \cite{AAHPT1, AAHPT2, AAHPT3, AAHPT4}. Critical phases are vital for understanding the transitions from localized to extended states, showing a range of fascinating phenomena including dynamical evolutions \cite{dynamical_evolution1, dynamical_evolution2, dynamical_evolution3}, critical spectral behavior \cite{critical_behavior1, critical_behavior2, critical_behavior3, critical_behavior4}, and the multifractal nature of wave functions \cite{multifractal_nature1, multifractal_nature2, multifractal_nature3, multifractal_nature4}. Expanding upon the AAH model, variations incorporating different forms of quasiperiodic disorder and interactions give rise to exotic phases, including critically localized states \cite{critically_localized_states1, critically_localized_states2, critically_localized_states3} and many-body localization \cite{many-body_localization1, many-body_localization2, many-body_localization3}. Recent research on non-Hermitian extensions of the one-dimensional AAH model has uncovered such a multicritical point marking the transition from localized to extended states, accompanied by PT symmetry breaking and topological phase transitions \cite{multicritical1, multicritical2, multicritical3}. However, the investigation of the interplay between the two-dimensional (2D) chiral topological edge modes and the non-Hermitian quasicrystal dissipation edge has not been previously explored.

In this context, our study not only reveals a complex phase diagram, but also establishes a profound relationship between the size of the non-Hermitian quasicrystal, the presence of imaginary zeros, and the increasing occurrence of first-order structural phase transitions. In the phase diagram, there are three distinct phases separated by two-phase boundaries: the extended phase (I), the localized phase (III), and the critical phase (II), as illustrated in Fig. \ref{fig:paradiagram}. Additionally, we have also discovered an increasing number of phase transitions (NPT) with the enlargement of the non-Hermitian quasicrystal size. We explain these first-order structure phase transitions (FOSPT) in the picture of phase splitting driven by imaginary zeros. Firstly, the quasi-periodic modulated potential contains certain points where the potential becomes zero. As the system parameters increase, a FOSPT occurs \cite{first-order_pt1, first-order_pt2}. Secondly, as the size of the system grows, the number of points with zero quasiperiodic imaginary potential increases. These zero points divide the imaginary potential into distinct domains, each having different positions for undergoing phase transitions as parameters vary. Consequently, the NPT increases with the size of the system. We have found that the NPT is equivalent to the number of zero points, which also matches the count of non-Hermitian domains. This phenomenon is unique to non-Hermitian systems and is absent in their Hermitian counterparts.

In Sec. \ref{model and phase diagram}, we establish the phase diagram by the inverse participation rate and provide an interpretation in terms of the effective low-energy non-Hermitian model. In Sec. \ref{size analysis}, we explore the relationship between the first-order structure phase transition and the dimensions of the system. Section \ref{Conclusion} is devoted to our conclusion.
\section{model and phase diagram}
\label{model and phase diagram}
\begin{figure}[]
	\centering
	\includegraphics[width=3.5in]{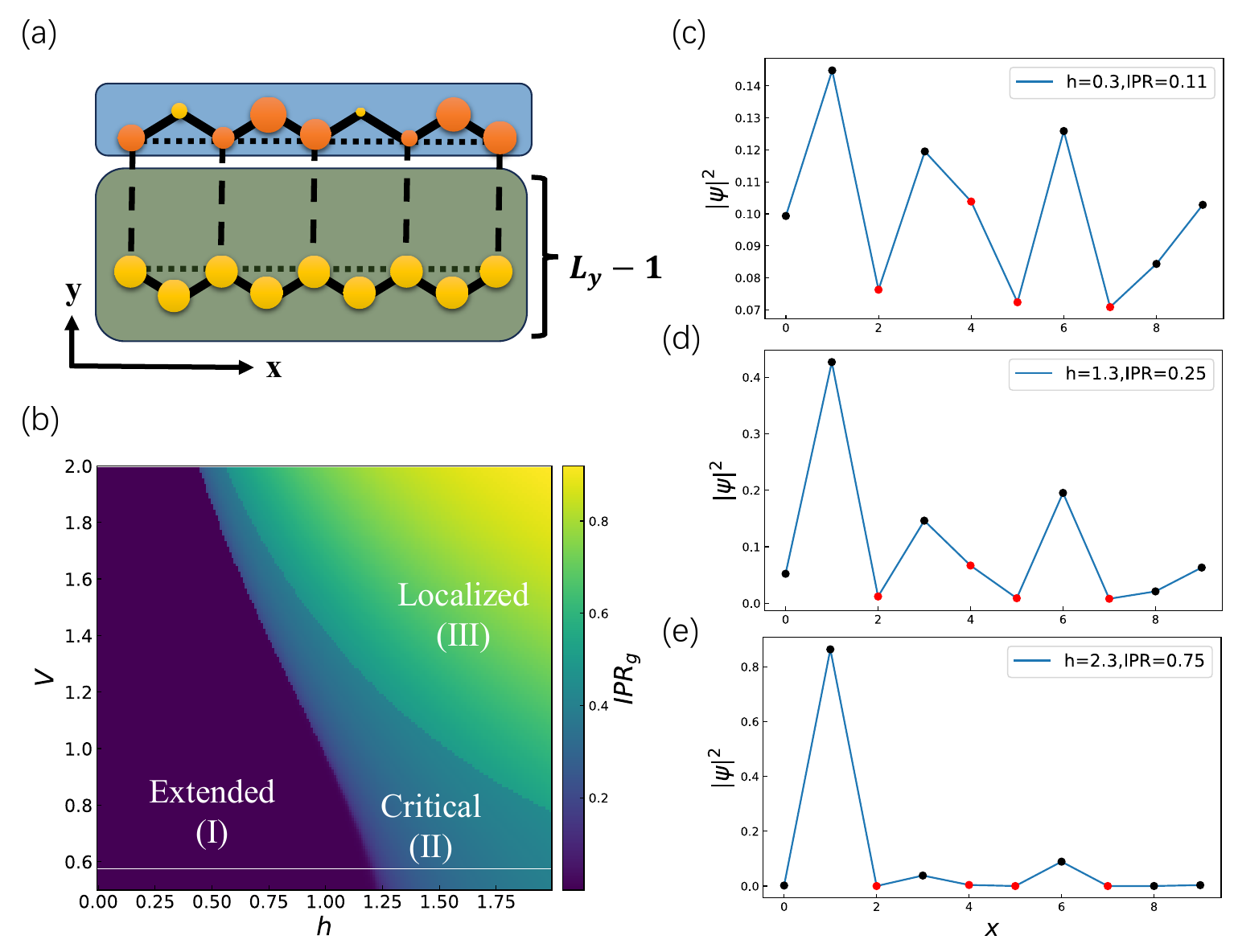}
	\caption{(a)-(e) xPBC/yOBC. (a) Schematic of the 2D Haldane model with quasicrystal imaginary potential at the upper boundary. Quasiperiodic and zero potentials are denoted by orange/yellow spheres against blue/green backgrounds, respectively. Black solid lines represent NN hopping, black dotted lines represent NNN hopping, and black dashed lines indicate intermediate hidden layers ($L_y-1$). (b) The inverse participation ratio as a function of $V$ and $h$, revealing three phases separated by two critical lines: phase (\Rmnum{1}) with extended wave functions, PT-restored phase (\Rmnum{3}) featuring spatially localized wave functions, and critical phase (\Rmnum{2}) with fractional wave functions. Parameters: $L_x=20$, $L_y=20$. (c)-(e) The density $|\psi|^{2}$ as functions of $x$ at three distinct points, $h=0.3, 1.3, 2.3$, with $V=1$, following Eq. (\ref{eq:solution_low energy}). (c) IPR=0.11, (d) IPR=0.25, (e) IPR=0.75.}
\label{fig:paradiagram}
\end{figure}

The foundation of our study lies in the Hamiltonian, which exhibits different forms under varying conditions:
\begin{equation}
H= \begin{cases}  H_{\text{AAH}}, & L_y = 1 \\ \text{two-chains}, & L_y=2, \\ H_{\text{Haldane}} + H_{\text{AAH}}^{\text{edge}}, & L_y \rightarrow \infty, \end{cases}
\label{eq:whole Hamiltonian}
\end{equation}
with two critical points of longitudinal dimensions $L_y= 1, 2$. When $L_y=1$, the system returns back to the 1D non-Hermitian AAH model which plays a key role in this paper
\begin{equation}
H_{\text{edge}}^{\text{AAH}} = \sum_{n} V\cos (2 \pi \alpha n + i h) c_n ^{\dagger} c_n,
\label{eq: quasi_potential}
\end{equation}
where $c_{n}^{\dagger}$ and $c_{n}$ are the creation and annihilation operators for a particle at the $n$-th site. $V$ and $h$ are the amplitude and imaginary phase of the potential, and $\alpha$ is an irrational number for a QC. Throughout this paper, we set $V=1$ as the energy unit. 

Since $\alpha$ is an irrational number, it can be approximated by a sequence of rational numbers $p_n/q_n$, where $p_n$, $q_n$ are prime numbers and $p_{n}$, $q_{n} \rightarrow \infty$ as $n \rightarrow \infty$. In numerical simulations, it is common practice to consider a finite (yet arbitrarily large) number of sites $L = q_n$ on a ring with periodic boundary conditions, where the occupation amplitudes are $\psi_{n+L} = \psi_n$.

When $L_y=2$, the system is no more than one-dimension, but a two-chain system with a 1D AAH chain coupled to a hopping-only chain \cite{critical_behavior4, transfer_matrix1}. Here we focus on the non-Hermitian Haldane model with quasiperiodic complex potential $H^{\text{AAH}}_{\text{edge}}$ on the upper boundary [Fig. \ref{fig:paradiagram}(a)], where height and circumference are $L_y$ and $L_x$. In the limits $L_y \rightarrow \infty$, the Hamiltonian $H=H_{\text{Haldane}} + H_{\text{edge}}^{\text{AAH}}$ and the Haldane model is an important model for describing the topological insulator \cite{Haldane1,Haldane2},
\begin{equation}
H_{\text{Haldane}}=t_1 \sum_{\langle n m\rangle} c_n^{\dagger} c_m+t_2 \sum_{\langle\langle n m\rangle\rangle} e^{i \phi_{n m}} c_n^{\dagger} c_m,
\label{eq: Haldane}
\end{equation}
where the nearest-neighbor (NN) couplings are denoted by $t_{1} = 1$, and the next-nearest-neighbor (NNN) coupling coefficients are $t_{2} e^{i \phi_{n m}}$ with amplitude $t_{2}$ and phase $\phi_{n m}$. The symbols $\left\langle n, m \right\rangle $ and $\left\langle \left\langle n,m \right\rangle \right\rangle $ denote the NN and NNN hopping, shown in Fig. \ref{fig:paradiagram}(a) as black solid and black dotted lines, respectively. The complex phase $e^{i \phi_{n m}}$ accounts for the NNN hopping, and we set the positive phase direction to be clockwise ($| \phi_{n m} = \frac{\pi}{2}|$). Below we consider both x direction as a periodic boundary condition (PBC) and y direction as an open boundary condition (OBC), i.e., a cylindrical geometry.

Our focus is on revealing the critical phase emerging in the non-Hermitian quasiperiodic boundary, which offers valuable insights into the boundary effects in dissipative systems. To determine the phase diagram of the Hamiltonian (\ref{eq:whole Hamiltonian}) under $L_y \rightarrow \infty$ condition, we compute the inverse of the participation ratio (IPR)
\begin{align}
\text{IPR}_n = \frac{\sum_m \left|\braket{\psi_{n,m}^R|\psi_{n,m}^R}\right|^4}{\left|\sum_m \braket{\psi_{n,m}^R|\psi_{n,m}^R}\right|^2},
\end{align}
as a function of $V$ and $h$, shown in Fig. \ref{fig:paradiagram}(b). Here, $\ket{\psi_{n, m}^{R}}$ represents the right eigenstate of the $H$ corresponding to the energy eigenvalue $E_n$, and $m= 1,\dots, 2L_x$. Specifically, when $n$ corresponds to the ground state (denoted as $g$), $\ket{\psi_g}$ represents the ground state, and $\text{IPR}_g$ quantifies the localization of the ground state.
Phase (I) with delocalized states has an $\text{IPR}_g\simeq1/L\simeq 0$, the PT-restore phase (III) with fully localized states, on the other hand, has an $\text{IPR}_g\simeq1$, and the critical phase (II) with fractional states fall in between, with IPR$_g$ values ranging from 0 to 1, indicating an intermediate level of localization, as shown in Fig. \ref{fig:paradiagram}(b).

In contrast to the one-dimensional case, where only a transition from the extended phase to the localized phase is observed \cite{multicritical1, multicritical2, multicritical3}, and distinct from the mobility edge resulting from the coupling of two chains \cite{multifractal_nature4}, our scenario gives rise to a unique critical phase. Within this context, FOSPT occurs in the ground state [see Appendix \ref{first-order phase transition}]. In the following Appendix \ref{fractional dimension, symmetry breaking, and fidelity}, we will present a more comprehensive analysis of the complete phase diagram and offer analytically derived phase boundaries using an effective model.

For a 2D chiral topological insulator, chiral modes can only exist on the boundary of topological materials. In the continuous limit, the effective Hamiltonian in the low-energy is described as $H_{chiral}=v_{f} k$, where $v_{f}=\frac{ \partial H_{chiral} }{ \partial k }_{k=k_{f}}$ is fermi velocity, and $k$ is the vector of the chiral modes. In the long-wavelength and low-frequency regimes, excitations are restricted to one-direction propagation and protected by non-trivial bulk topology.

Then, we consider the effects of dissipation which the boundary potential has a non-zero imaginary part, and the effective low-energy Hamiltonian \cite{hybrid_skin-topological_effect1, hybrid_skin-topological_effect2, hybrid_skin-topological_effect3} reads as 
\begin{equation}
H_{chiral}=v_{f} k + i V
\label{eq: low Hamiltonian}
\end{equation}
where the imaginary part of eigenvalues is dependent on the on-site dissipation potential $V$. The Schr\"{o}dinger equation of the dissipation chiral modes is
\begin{equation}
\left[-i v_f \frac{d}{d x}+i V(x)\right] \psi(x)=(\epsilon_{r}+i \epsilon_{i})  \psi(x),
\label{eq:low energy}
\end{equation}
where $\epsilon_{i}$ is the imaginary part of the eigenenergy.

Then, we get the solution of Eq. (\ref{eq:low energy})
\begin{equation}
\psi(x)=\frac{1}{\sqrt{C}} \exp \left(i \frac{\epsilon_{r} }{v_f} x\right) \exp \left(\int_0^{ L_x } d x^{\prime} \frac{ V \left(x^{\prime}\right)-\epsilon_{i}}{v_f}\right),
\label{eq:solution_low energy}
\end{equation}
where $1/\sqrt{C}$ is the normalization factor and the integration region (0, $L_x$) is on the dissipation boundary.

Since the 2D Haldane model is periodic in x, the periodic boundary condition gives $\psi( L_x ) = \psi(0)$. Then, we have
\begin{equation}
i \frac{ \epsilon_r }{v_f} L_x-\frac{\epsilon_i }{v_f} L_x+\frac{1}{v_f} \int_0^{L_x} d x^{\prime} V \left(x^{\prime}\right)=2 i \pi n,
\label{eq: periodic condition}
\end{equation}
where $n \in \mathbb{Z}$. Then, the Eq. (\ref{eq: periodic condition}) can be reduced to 
\begin{align}
\epsilon_r & =\frac{1}{v_f} \frac{2 \pi n}{L_x}, \\
\epsilon_i & =\frac{1}{L_x} \int_0^{L_x} d x^{\prime} V \left(x^{\prime}\right) = \widetilde{V},
\label{eq: epsilon_i}
\end{align}
where the imaginary part of eigenenergy $\epsilon_i$ is the average value of imaginary potential $\widetilde{V}$.

The first two components (\ref{eq:solution_low energy}), similar to plane waves $exp(i k x)$, are uniformly distributed throughout the entire space. However, when the sign of $v_f$ is fixed, the third component (\ref{eq:solution_low energy}) introduces exponential growth or decay is possible, depending on the sign of $sgn(V \left( x \right)-\epsilon_{i})$ (either 1 or -1). For example, at position $x_c$, a change in the sign of the imaginary potential occurs like a step function, specifically when $V(x<x_c)<0$ and $V(x>x_c)>0$. If $v_f<0$, the edge state will exhibit a peak at $x_c$. Conversely, if $V(x<x_c)>0$ and $V(x>x_c)<0$, and $v_f>0$, the edge wave function will also display a peak at $x_c$. These are exactly the black circles in Figs. \ref{fig:paradiagram}(c)-(e), which are the positive imaginary potentials.

In the specific case of our study, the dissipative potential takes the form of a quasi-periodic pattern, as shown in Eq. (\ref{eq: quasi_potential}). Substituting Eq. (\ref{eq: quasi_potential}) into Eq. (\ref{eq: epsilon_i}), it can be reduced to
\begin{equation}
\epsilon_i= -\frac{V \sin ^2(\pi  \alpha L_x)}{\pi  \alpha L_x} \sinh (h).
\end{equation}
Consequently, according to the above discussion, there will be a peak approximating a period, due to the quasi-periodic nature of the imaginary potential,  as depicted in Figs. \ref{fig:paradiagram}(c)-(e). To better visualize the phase diagram [Fig. \ref{fig:paradiagram}(b)] in different phases, we selected three positions along the $V = 1$ line in Fig. \ref{fig:paradiagram}(b): $h = 0.3, 1.3$, 2.3. Among these positions, two critical points are evident: the extended-critical transition point at $h_1 = 0.97$ and the critical-localized phase transition at $h_2 = 1.41$. 

The red and black circles represent positive and negative on-site potentials, respectively, with a system size of $L_x = 10$. As shown in Fig. \ref{fig:paradiagram}(c), when the imaginary phase $h = 0.3$, the overall density of the wave function exhibits minor fluctuations along the x-direction, corresponding to IPR$=0.11\approx1/L_x$.  At this point, the quasi-periodic potential is weak enough to be seen as a perturbation, $V$ can be seen as part of the elliptical complex energy spectrum, as shown in Appendix \ref{Dissipation in two limit cases:}. As the imaginary potential gradually increases $(h = 1.3)$, certain randomly distributed positions experience higher density, while others exhibit decreased density. This leads to an intermediate value IPR$=0.25$, indicating a wave function reminiscent of a fractal-like state, as shown in Fig. \ref{fig:paradiagram}(d). Finally, for a larger imaginary potential $h = 2.3$, the wave function localizes randomly at any position, with the maximum occupation $|\psi|^2_{max}\approx0.9$ is observed at $L_x=1$ (IPR=0.75), and the sharp peaks occur precisely at the transitions from black (positive) to red (negative), as shown in Fig. \ref{fig:paradiagram}(e).

\section{first-order structure phase transition}
\label{size analysis}
\begin{figure}[]
	\centering
	\includegraphics[width=3.6in]{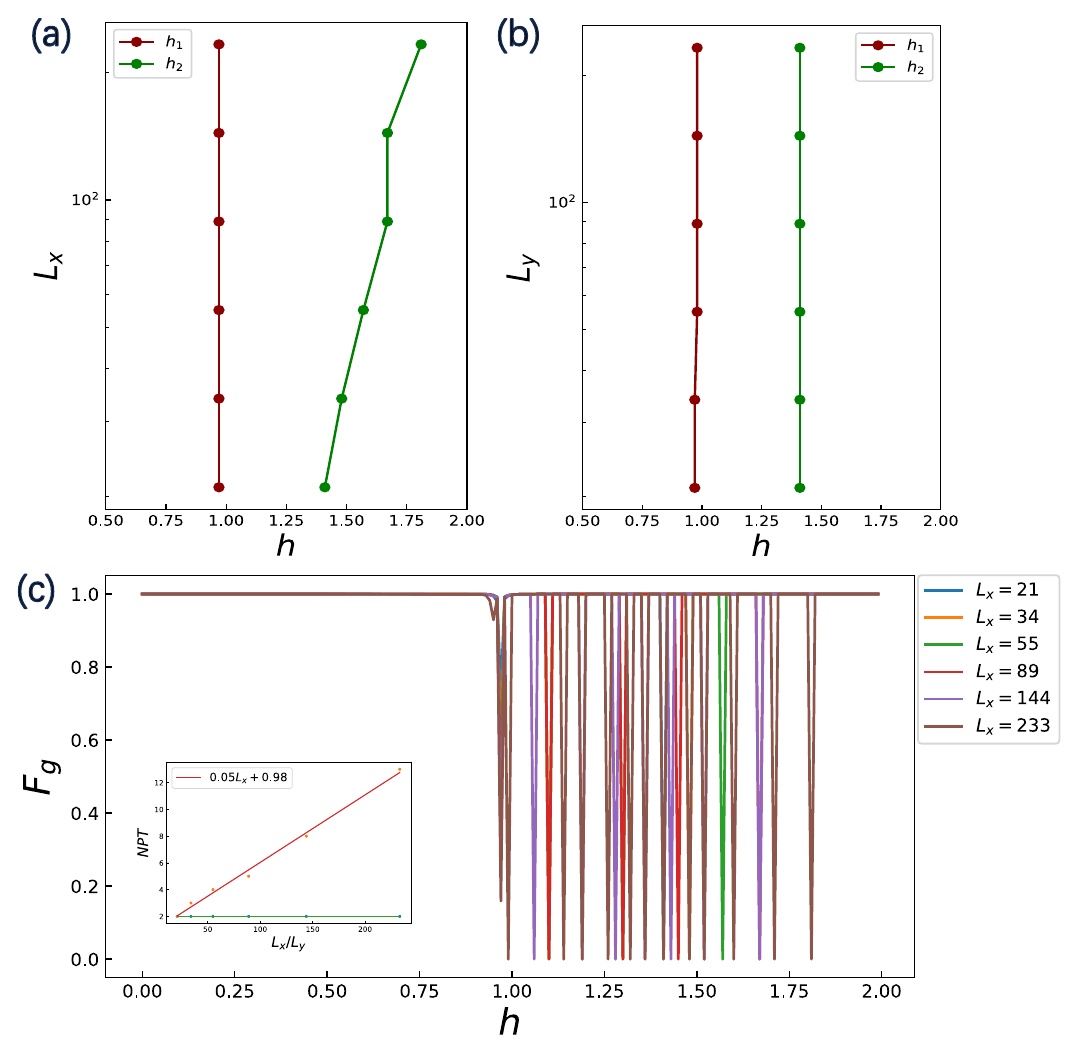}													
	\caption{(a) The transverse dimension $L_x$ is a function of $h_1/h_2$. (b) The longitudinal dimension $L_y$ is a function of $h_{1}/h_2$. (c) The ground fidelity $F_g$ plotted against $h$ with $L_x=21, 34, 55, 89,144, 233$. Inset: The NPT as a function of $L_x/L_y$. The red line represents $L_x$, and the fit yields $\text{NPT}=0.05 L_x +0.98 $. The green line corresponds to $L_y$.}
\label{fig: size_analysis}
\end{figure}
Here, we explore how the extended-critical and the critical-localized phase transition points $h_1, h_2$ are also affected by the dimensions $L_x, L_y$. The $log(L_x)$ as a function of $h_1$ is shown in Fig. \ref{fig: size_analysis}(a), where we can see $h_1\approx 0.97$ is almost a constant and $log (L_x) = k h_2 + c$ is a linear function, where $k$ and $c$ are constants. As the size of the system increases, the alternation of positive and negative imaginary potentials becomes more pronounced. It is due to the effect of positive imaginary potential gain that the wave function (Eq. (\ref{eq:solution_low energy})) tends to disperse more to locations marked by positive imaginary potentials as the system dimension expands. This inherent tendency leads to an increase in the localization transition points of the wave function as the system dimensions ($L_x$) increase. In turn, $h_1, h_2$ is shown in Fig. \ref{fig: size_analysis}(b) as $L_y$ varies, when $h_1$, $h_2$ are all essentially constants.

We also find another interesting phenomenon that the critical phase fragments into multiple parts, which cannot happen in the Hermitian system. We have plotted the fidelity of the ground state $F_g$ as a function of $h$ for different sizes $L_x=21, 34, 55, 89, 144, 233$ in Fig. \ref{fig: size_analysis}(c). It can be seen that the NPT increases as $L_x$ increases. To know the relationship between NPT and system size, we plot NPT as a function of $L_x/L_y$, as shown in the subplot of Fig. \ref{fig: size_analysis} (c). It can be seen that when $L_y$ increases, the NPT is kept constant because there is no increase in the size of $V (x)$, as shown by the green line. However, the red line represents the NPT vs $L_x$. The result given by the linear fit is NPT $= 0.05L_x + 0.98$. 0.05 means that for every 20 lattices, there is a zero point increase in the imaginary potential, and 0.98 means that there is a phase transition from a delocalized phase to a localized phase at the beginning. Then for the $V(x)$ as in Eq. (\ref{eq: quasi_potential}), there will be sites where $Im(V(x)) \approx 0$, which is where the FOSPT occurs with parameters change. As the $L_x$ increases, the number of zero imaginary potential points increases. These zeros partition the potential into distinct domains, each hosting different phase transition points as the parameters vary. This is the reason why the NPT increases as the $length(V(x))$ increases.

\begin{figure*}[]
	\centering
	\includegraphics[width=7.2in]{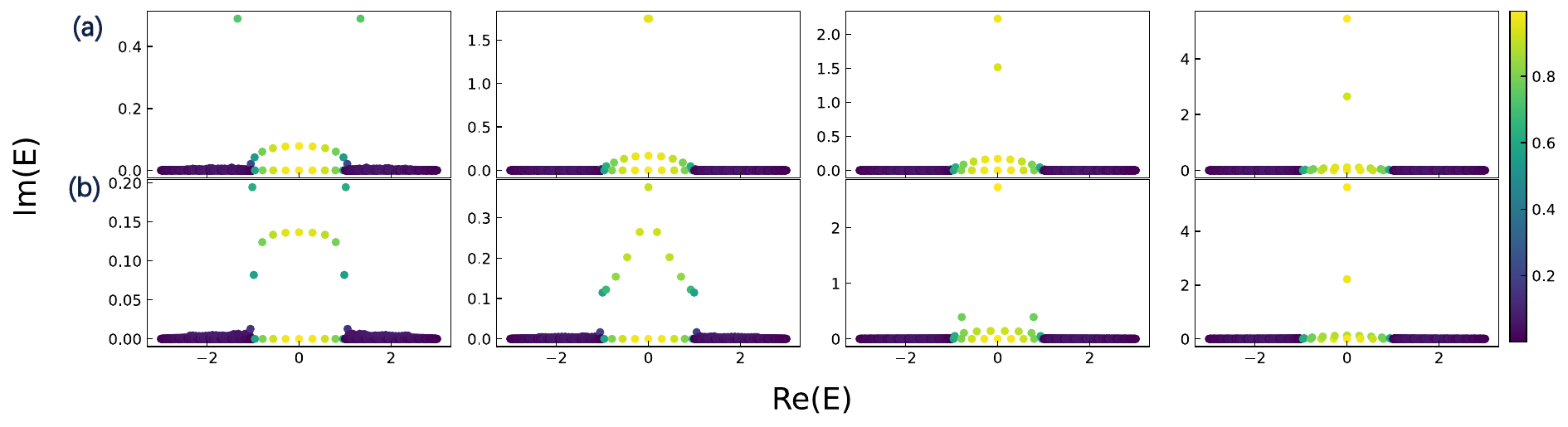}													
	\caption{Energy spectrum of the Haldane model with different imaginary potential impurities. (a) The energy spectra with two nearest-neighbor imaginary impurities for $\gamma=1.600$, 3.356, 3.500, 6.000 from left to right. (b) The energy spectra with two next nearest-neighbor imaginary impurities for $\gamma=1.600$, 2.000, 3.400, 6.000 from left to right. The parameters are chosen as $L_x=20$, $L_y=20$.}
\label{fig: phase transition in imaginary potential}
\end{figure*}
In this following, we aim to establish the connection between the FOSPT and the imaginary potential. Fig. \ref{fig: phase transition in imaginary potential} illustrates the complex energy spectrum of two scenarios: one with nearest-neighbor imaginary potential [Fig. \ref{fig: phase transition in imaginary potential}(a)], defined as
\begin{equation}
H_{NN}=H_{Hermitian} + i V_1 + i V_2 /2,
\label{eq: HNN}
\end{equation}
and another with next-nearest-neighbor imaginary potential [Fig. \ref{fig: phase transition in imaginary potential}(b)] given by
\begin{equation}
H_{NNN}=H_{Hermitian} + i V_1 + i V_3 /2,
\end{equation}
where $H_{Hermitian}$ is an arbitrary Hermitian matrix, we consider the 2D Haldane model $H_{Hermitian}=H_{Haldane}$ ( the simplest PT-symmetric matrix is presented in Appendix \ref{PT phase transition}). In $V_i$, ``$i=1, 2, 3$'' represent the on-site potential of any position on the boundary and are increasing in order. The color of points represents the density of edge $|\psi_{edge}(h) |^{2}$.

In the first plot with $V=1.600$, points close to the x-axis depict topological boundary states influenced by the imaginary potentials. Their energy spectra form a semicircle with the $x$-axis, exhibiting a skinning effect, as shown in Fig. \ref{fig: phase transition in imaginary potential}(a). The two points with the largest imaginary part correspond to two non-topological boundary states that are progressively confined towards the boundary due to the impact of the imaginary potential. The colors in the figure represent the density at the boundary, revealing that the topological boundary states primarily occupy the boundary, while states outside the energy gap are bulk states, except for the non-topological boundary states. At $V=V_c=3.356$, a critical phase transition occurs the two non-topological boundary states are degenerate. If the imaginary potential is slightly larger, i.e., $V=3.500$, the real part of these two states remains the same, but the imaginary part differs, resulting in a PT phase transition. With further increases in $V$, the energy of the non-topological boundary states increases in tandem with the imaginary potential, and the imaginary parts of the topological boundary states become nearly zero, restoring PT symmetry.

We also consider the case of next-nearest-neighbor imaginary potentials for $V = 1.600, 2.000, 3.400, 6.000$. The energy spectra are more or less the same at the beginning, and the energy of the two non-topological boundary states undergoes a degenerate at $V = V_{c1} = 2.000$, which is also the first critical point. However, when $V=V_{c2}=3.400$, two additional non-topological boundary states emerge near the topological boundary states. These two non-topological boundary states undergo a second PT phase transition. As in Fig. \ref{fig: phase transition in imaginary potential}(b), after the occurrence of two PT phase transitions, there are only non-topological boundary states localized at the two imaginary potential points, as in Fig. \ref{fig: phase transition in imaginary potential}(b). This observation from the simplest case of two imaginary potentials can be extended to scenarios involving multiple imaginary potentials, where PT phase transitions occur whenever non-adjacent imaginary potentials are present, see Fig. \ref{fig: size_analysis}(c).

If these zeros are replaced by finite imaginary potentials, we find that the NPT returns to the situation in Fig. \ref{fig:paradiagram}(b) and does not vary with $L_x$, as shown by the green line. Moreover, when the length of the imaginary potential is constant, i.e. length($V(x)$) = 20, we plot the NPT as a function of $L_y$, again still with the green line. This means the phase transition depends only on the imaginary potential change.

\section{Conclusion}
\label{Conclusion}
We uncover the rich phase diagram of the two-dimensional Haldane model with edge quasi-periodic dissipation. Overall, the fractional dimension, the largest imaginary part of the eigenvalues, the scaling exponent, and the ground state fidelity provide valuable insights into the localization properties and phase transitions in the system. The system exhibits extended, critical, and localized phases, with the criticality appearing between the extended-critical and critical-localized transitions. In the low-energy approximation, we show phase transitions of the wave function in different phases. We then projected the original Hamiltonian onto the boundary subspace and obtained an effective two-chain model which yielded a phase diagram similar to that of the original Hamiltonian, effectively capturing its phase transition properties.

We also analyzed the effect of transverse and longitudinal dimensions on the phase transition. The results show that the longitudinal dimensions do not affect the phase transition, but the critical point from the critical phase to the localized phase becomes larger with increasing transverse dimensions, which is due to the gain effect of the positive imaginary potential, which weakens the localization of the wave function. Then we also find the phenomenon of critical phase tearing, which is unique in non-Hermitian systems, due to the first-order structural phase transition of the system caused by the zeros of the quasi-periodic imaginary potential.

\begin{acknowledgments}
We are grateful to Gao Xianlong, Yiling Zhang, Xin-Ran Ma, Qian Du, Yufei Zhu for valuable suggestions on the manuscript. This work is supported by NSFC Grants No. 11974053 and No. 12174030.
\end{acknowledgments}

\appendix
\section{FIRST-ORDER PHASE TRANSITION}
\label{first-order phase transition}
\begin{figure}[]
	\centering
	\includegraphics[width=3.5in]{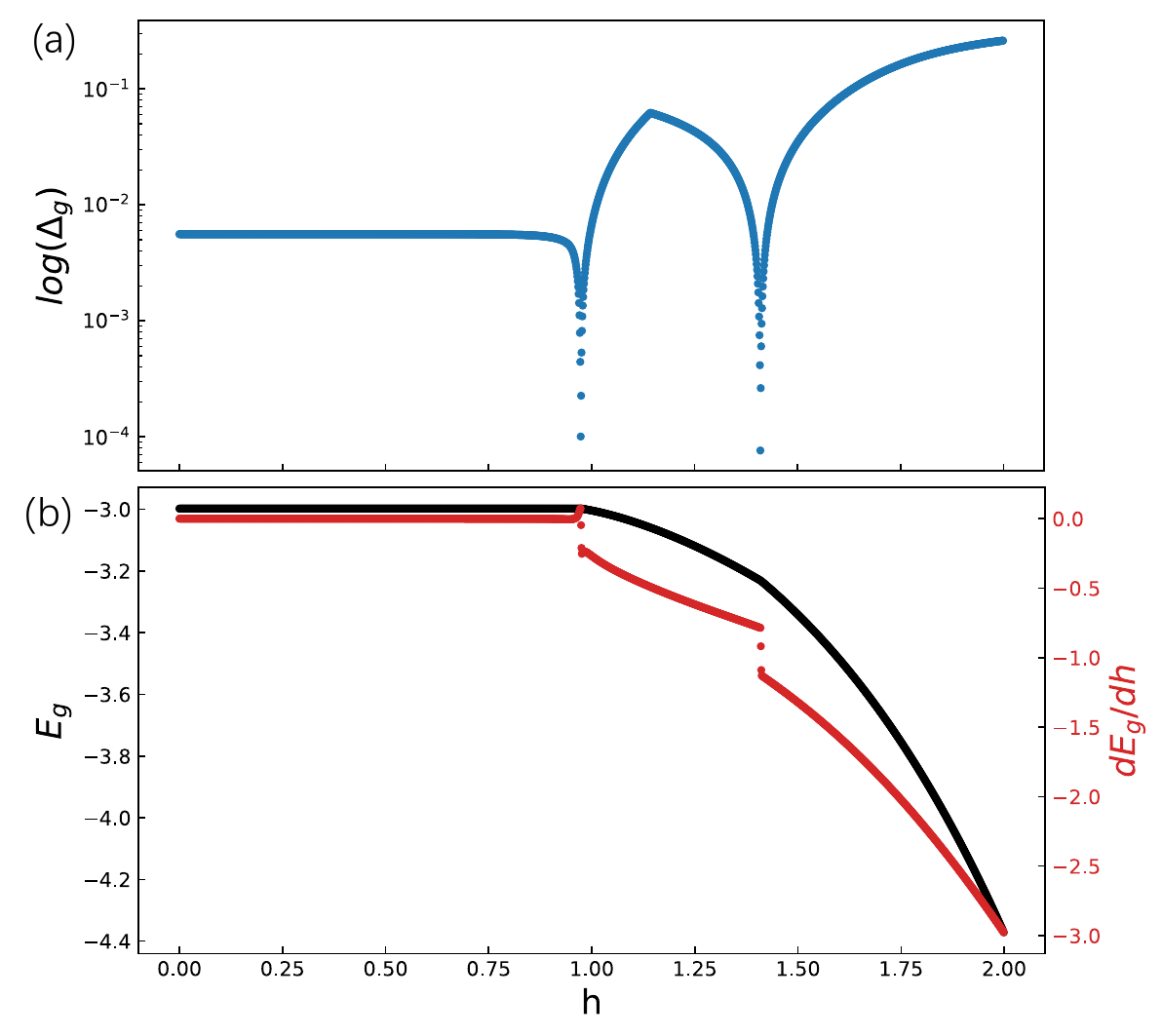}
	\caption{(a) The logarithm of energy gap $log(\Delta_{g})$ as a function of $h$. (b) The first-order derivative of the ground state energy $dE_{g}/dh$ as a function of $h$. Both $log(\Delta_{g})$ and $dE_{g}/dh$ exhibit discontinuities at $h_1$ and $h_2$. The parameters are chosen as $L_x=20$, $L_y=20$.}
\label{fig: energy}
\end{figure}
The quantum phase transition in the Hermitian system refers to the nonanalyticity of avoided level-crossing or actual level-crossing \cite{quantum_phase_transition}. However, the spectra of the non-Hermitian system are generally complex. There are some line gaps and point gaps that have been found in the non-Hermitian AA systems \cite{ non-Hermitian6}. When the parameters are changed, the line gaps may be experienced many times close to reopening processes if there is more than one zero point. Traditionally, quantum phase transition is characterized by singularities of the ground state energy and the energy gap between the ground state and the first excited state. First-order quantum phase transition is identified by abrupt changes in the first derivative of the energy. 

In Fig. \ref{fig: energy} (a), we plot the logarithm of the energy gap $log(\Delta_g)$ as a function of $h$, where
\begin{align}
\Delta_{g} =E_{f} - E_{g}.
\end{align}
$E_f$ is the first excited state energy and $E_g$ is the ground state energy. There are two discontinuous points $h_1$ and $h_2$, at which the gap is closed. It also corresponds to the phase transition in Fig. \ref{fig:paradiagram}(b).

To determine the phase transition type, we calculated the ground state energy $E_g$ and its first derivative $dEg /dh$ as a function of $h$. In Fig. \ref{fig: energy}(b), the black dots represent the ground state energy, while the red dots represent its first derivative. Although $E_g$ is continuous, the discontinuity of $dEg /dh$ at $h_1$ and $h_2$ shows the first-order nature of quantum phase transition, similar to in Fig. \ref{fig: energy}(a).

\section{FRACTION DIMENSION, PT-SYMMETRY BREAKING, AND FIDELITY}
\label{fractional dimension, symmetry breaking, and fidelity}
\begin{figure*}[]
	\centering
	\includegraphics[width=7in]{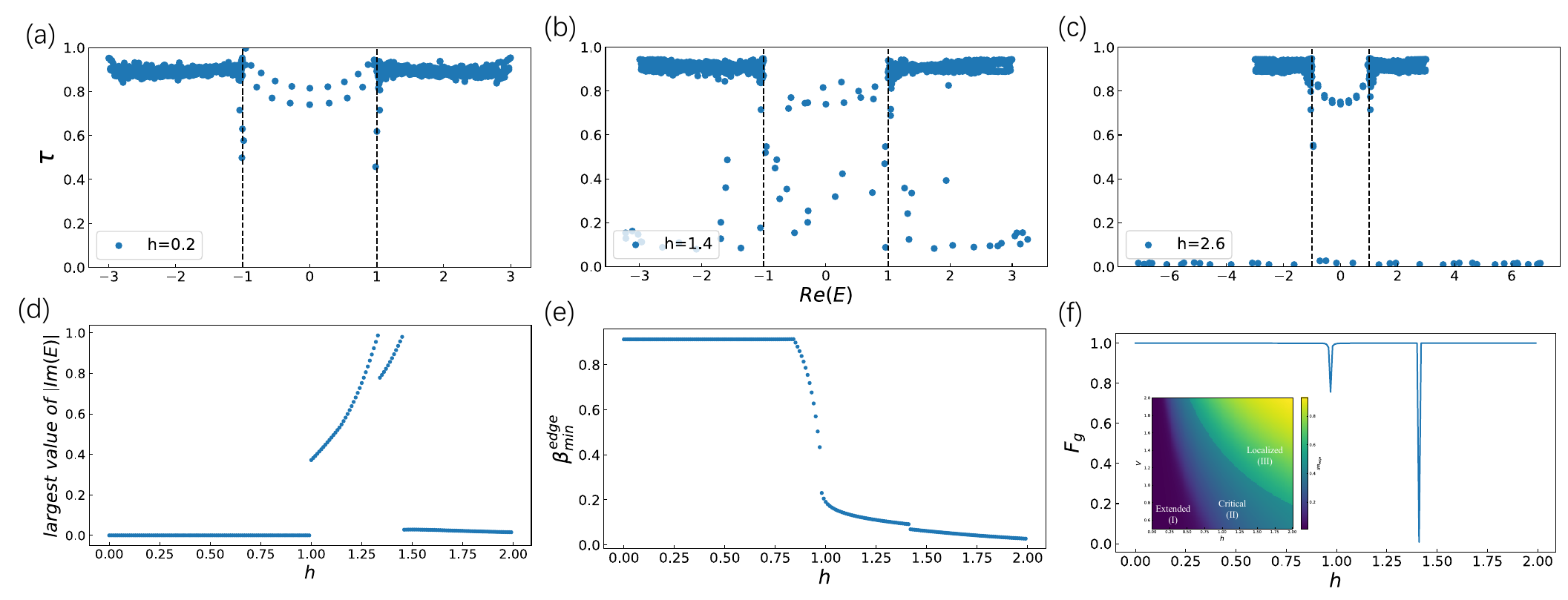}													
	\caption{(a)-(c). The fractal dimensions $\tau$ as a function of the real part of eigenenergies $Re(E)$ for three different $h=0.2, 1.4, 2.6$. In (d), the largest value of $|Im(E)|$ is a function of $h$. (e) The minimal scaling exponent as a function of $h$. (f) The fidelity of ground states $F_g$ versus $h$. Abrupt changes at $h_1$ and $h_2$ are present in (d)-(f). Inset: Phase diagram of the effective two-chain model $H_{eff}$. The parameters are chosen the same as Fig. \ref{fig: phase transition in imaginary potential}.}
\label{fig:edge_fidelity}
\end{figure*}
The localization behavior of the system's wave function is a crucial observable that requires precise measurements. Wave functions are commonly characterized by their fractal dimension, quantified by the IPR, which follows a scaling relation of 
\begin{align}
\text{IPR} \sim (L_x)^{-\tau},
\end{align}
 where $\tau$ represents the fractal dimension (FD). The FD provides a valuable perspective to understand how states expand and fluctuate as the system size increases. Similar to IPR, When $\lim_{L_x\rightarrow\infty}\tau=1$, it indicates an extended wave function. Conversely, if the wave function is localized with peaks only at a few lattice points and negligible amplitudes elsewhere, it implies $\lim_{L_x\rightarrow\infty}\tau=0$. Fractal wave functions exhibit FD values within the range of $0<\lim_{L_x\rightarrow\infty}\tau<1$.

In Figs. \ref{fig:edge_fidelity}(a-c), we compare the fractal dimensions of the extended, critical, and localized phases. Specifically, we examine three points along the $V=1$ line in Fig. \ref{fig:paradiagram}(b): $h=0.2, 1.4,$ and $2.6$. The two black dashed lines in all plots correspond to $Re(E)=\pm1$ positions, between which states represent topological boundary states when $h$ is small. In Fig. \ref{fig:edge_fidelity}(a), the majority of states are concentrated at the top, indicating extended states with $\tau=1$. Fig. \ref{fig:edge_fidelity}(b) shows a decreasing trend in both topological and non-topological boundary states, with $\tau$ values ranging from $0.1$ to $0.6$. These non-topological edge states refer to those influenced by the presence of an imaginary potential, which tends to localize at the boundary. Furthermore, it is worth noting that at $\tau=0.6$, we observe the separation between extended states and fractal states localized at the dissipative boundary. 

Under a strong imaginary potential at $h=2.6$ in Fig. \ref{fig:edge_fidelity}(c), we discover that all non-topological boundary states move to the bottom with $\tau=0$, and their count precisely matches the number of topological boundary states, which is $L_x$. However, the topological boundary states return to the top of the plot, indicating their return to extended states unaffected by the imaginary potential.

In fact, the impact of the imaginary potential on the system's topological states is observed in the critical phase, as depicted in Figs. \ref{fig:edge_fidelity}(c). A captivating question arises: How does the variation of the parameter influence the imaginary part of the topological boundary states? Fig. \ref{fig:edge_fidelity}(d) illustrates the behavior of the largest imaginary part $|Im(E)|$  vs $h$. Remarkably, a sudden surge from zero occurs at $h_1 \approx 0.97$, followed by an abrupt decline to zero at $h_2 \approx 1.41$. Remarkably, when $h<0.97$ (PT-symmetric) and $h>1.41$ (PT-restored), $max(Im(E_e))$ remains close to zero, in agreement with Figs. \ref{fig:edge_fidelity}(a) and (c). However, it is in the intermediate region that we observe the most pronounced impact of the imaginary potential, resulting in non-zero values of $Im(E_e)$. This observation verifies our earlier conjecture and highlights the intricate interplay between the parameter variation and the imaginary part of the topological boundary states, see Eq. (\ref{eq:low energy}).

One calculation similar to the fractal dimension is the scaling exponent, which can be obtained from the on-site probabilities of any wave function \(\psi_n\). The on-site probability is restricted to the boundary. According to the fractal theorem, the scaling of the maximum on-site probability is expressed as 
\begin{align}
max(p_{n,\text{edge}}) \sim (2L_x)^{-\beta_{n}^{\text{edge}}}, 
\end{align}
where $p_{n,edge} = |\psi_{n, \text{edge}}|^2 $ and $\text{edge}=1,\dots,2L_x$. To determine the extended, critical, and localized wave functions, we only need to investigate the minimum value of the exponent $\beta_{min}^{\text{edge}}$.

Referring to Fig. \ref{fig:edge_fidelity}(e), we focus on the boundary scaling exponent of the ground state. Strikingly, we observe a precipitous decline in $\beta_{min}^{edge}$ precisely at the critical points \(h_1\) and \(h_2\). When \(h < h_1\), the system manifests an extended phase, thereby approximating $\beta_{min}^{edge}$ to 1. Conversely, for \(h > h_2\), the system assumes phase (III), whereby the boundary wave functions localize, yielding $\beta_{min}^{edge}\approx0$. Within the critical phase, $\beta_{min}^{edge}$ spans the interval (0.1, 0.25), providing evidence of a fractal nature of the ground state.

We have carried out various calculations, some of which are based on knowledge of the physical properties of the system. However, in Fig. \ref{fig:edge_fidelity}(f), we investigate the behavior of the fidelity $F_{g}$ vs $h$. Fidelity has a distinct advantage in that it does not require prior familiarity with the order parameters or symmetries of the system. Typically, we can expect that as the ground state structure undergoes sharp changes, the fidelity will abruptly decrease near the critical points of the system. We focus on the boundary subspace and explore the boundary fidelity, quantified as:
\begin{align}
F_{g}(h,\delta h)=\left| \braket{ \psi_{g, \text{edge}}(h)| \psi_{g, \text{edge}} (h+\delta h)} \right|,
\end{align}
where $\delta h$ is a small quantity, $\ket{ \psi_{g, \text{edge}} (h) } = \sum_{edge} \ket{\psi_{edge} }\braket{ \psi_{edge}| \psi_{g}(h) }$ and $\ket{ \psi_{ g } (h) }$ satisfies the eigenvalue equation $H(h) \ket{\psi_{g} (h)} = E_{g} \ket{\psi_{g}(h) }$. At the critical points near $h_{1}$ and $h_{2}$, the overlap of the ground states $F_{g}$ undergoes a dramatic decrease, decreasing from $1$ to $0.76$ and from $1$ to $0$, respectively.

Based on the above description, the physical properties of the original Haldane model can essentially be captured by the characteristics of its boundaries. Therefore, we introduce the concept of the boundary effective Hamiltonian $H_{\text{edge}}$ to further comprehend the expanded-critical and critical-localized phase transitions depicted in Fig. \ref{fig:paradiagram}(b). Directly, we project the $H$ onto the boundary subspace using the boundary projection operator $P_{\text{edge}}$. The effective edge Hamiltonian is then given by:

\begin{align}
H_{\text{eff}} &= P_{edge}~H~P_{edge} \notag\\
		    &= H_{\text{AA}} + H_{\text{free}} + H_{c},
\end{align}
where $H_{\text{AA}} = \sum_m\left(a_m^\dagger a_{m+1} + \text{h.c.}\right) + V\cos(2\pi\alpha m + i h) a_m^\dagger a_m$ represents the non-Hermitian Aubry-André-Harper model, $H_{\text{free}} = \sum_{m} b_{m+1}^\dagger b_m + \text{h.c.}$ is the free chain with only the nearest-neighbor hopping term, and $H_c = a_m^\dagger b_m + \text{h.c.}$ represents their coupling, as seen in Eq. (\ref{eq:whole Hamiltonian}) with $L_y=2$. The $H_{\text{eff}}$ is reduced to a two-chain model 
\begin{equation}
\begin{aligned}
H_{ \text{eff} } = &\sum_{j=1}^2\sum_n \left[V_{j, m} c_{j, m}^{\dagger} c_{j, m}+t\left(c_{j, m}^{\dagger} c_{j, m+1}+\text { H.c. }\right)\right] \\
& +\lambda \sum_{m=odd} \left(c_{1, m}^{\dagger} c_{2, m}+\text { H.c. }\right).
\end{aligned}
\end{equation}
where $V_{1, m}=V \cos(2\pi\alpha m + i h)$ when $j=1$ and when $j=2$, $V_{2, m}=0$.

The phase diagram is almost unchanged, as shown in the subplot of Fig. \ref{fig:edge_fidelity}(f). In the subplot of Fig. \ref{fig:edge_fidelity}(f), two critical lines can also be seen dividing the whole phase diagram into three regions: extended, localized, and critical phase. This phase diagram is almost the same as Fig. \ref{fig:paradiagram}(c), except that the critical line from the extended phase to the critical phase is not particularly obvious, and the extended phase is also reduced. This is due to size effects, and in the case of $L_y \rightarrow \infty$ the subplot of Fig. \ref{fig:edge_fidelity}(f) will change back to Fig. \ref{fig:paradiagram}(c).

\section{DISSIPATION IN WEAK DISSIPATION CASE}
\label{Dissipation in two limit cases:}
\begin{figure}[]
	\centering
	\includegraphics[width=3.5in]{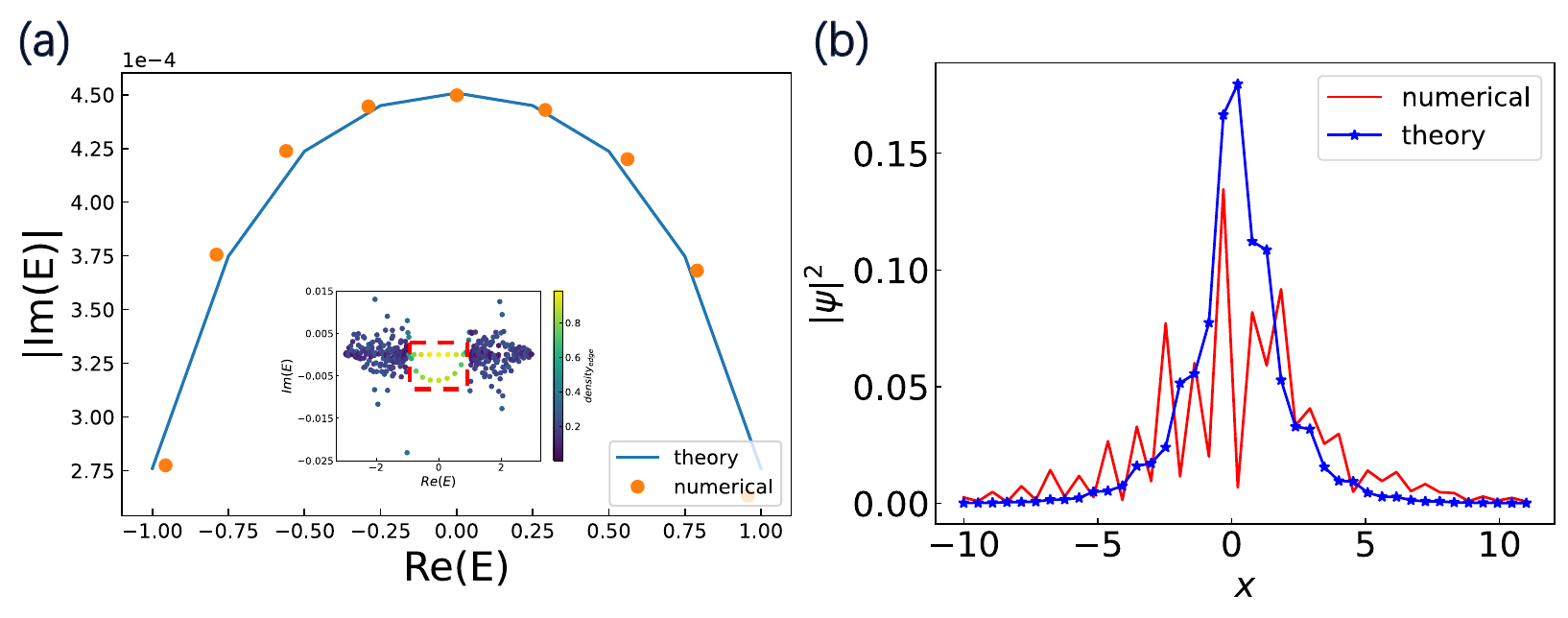}
	\caption{The phenomena in the regime of small imaginary potentials under gain/loss domain wall condition. (a) A comparison between numerical results (yellow dots) and theoretical predictions (blue line) for the complex energy spectrum of topological states. Inset: Energy spectra in whole complex energy space. (b) The topological protected edge state on gain/loss boundaries (red line) and the wave function for low-energy approximation. The parameters are chosen as $L_x=20$, $L_y=20$, $h=0.2$ and $\gamma=0.2$.}
\label{fig: weak}
\end{figure}
Energy bands that encircle a point gap with nonzero winding numbers under PBC will exhibit non-Hermitian skin effects under OBC. All eigenmodes are localized at the boundary. However, in the above cases, periodic boundary conditions ensure unidirectional chiral currents. Therefore, we introduce OBC at the gain-loss boundary, creating a dissipation domain wall. We define the Hamiltonian for the domain as 
\begin{align}
H_{domain}=H+\text{domain},
\label{eq:open Hamiltonian}
\end{align}
where $H$ is the Hamiltonian in Eq. (\ref{eq:whole Hamiltonian}) under $L_y \rightarrow \infty$, and $\text{domain}=H_G(x) + H_L(x)$, with subscripts $G$ and $L$ representing the '``gain'' and ``loss" domains, respectively. In each domain, gain ($x<0$) or loss ($x>0$) generates a constant on-site imaginary potential $\pm i \gamma$. This results in a purely imaginary shift in the spectra of $H$.

Fig. \ref{fig: weak}(a) shows the complex energy spectrum of this Hamiltonian $H_{domain}$ in the $|Re(E)|<1$ regime under weakly dissipative conditions.  where the vertical axis represents $\left| Im(E) \right|$. It corresponds to the region marked by the red dashed lines in the entire complex energy spectrum of the Hamiltonian, as shown in the subplots of Fig. \ref{fig: weak}(a). While the colors indicate the probability density of the eigenstates at the boundaries $density_{edge}= | \psi_{edge} |^{2}$. The complex energy satisfy half-ellipse equation, i.e., $\frac{\operatorname{Re}(E)^2}{a^2}+\frac{|\operatorname{Im}(E)|^2}{b^2}=1$.

In the limit of $h\rightarrow 0$, the quasiperiodic dissipation potential in Eq. (\ref{eq: quasi_potential}) undergoes a Taylor expansion
\begin{align}
\underset{h\rightarrow0}{\operatorname{lim}} V_n&=\underset{h\rightarrow0}{lim} \cos(2 \pi \alpha n+i h) \notag\\
						  &\approx \cos(2 \pi \alpha n)+ i \sin (2 \pi \alpha n) h +O(h^{2}).
\end{align}
The imaginary potential perturbs the edge modes, leading to a correction in their eigenenergies, given by
\begin{small}
\begin{align}
E^{1} &=\braket{ \psi^{0}_{e}| H_{edge} | \psi^{0}_{e} } \notag\\
	   &=\braket{ \psi^{0}_{e}| \cos(2 \pi \alpha n)+ i \sin (2 \pi \alpha n) h | \psi^{0}_{e} } \notag\\
	   &=\sum_{n=0}^{L_{x}} \left| \psi_{n}^{0} \right|^{2} \cos(2 \pi \alpha n) + ih \sum_{n=0}^{L_{x}} \left| \psi_{n}^{0} \right|^{2} \sin (2 \pi \alpha n) \notag\\
	   &=\operatorname{Re}(E^{1}) + i \operatorname{Im}(E^{1}),
\label{perturbation energy}
\end{align}
\end{small}
where $\ket{\psi_e^{0}}$ is an eigenfunction of $H$ satisfying $H \ket{\psi_e^{0}} =E^{0} \ket{\psi_e^{0}}$, $\operatorname{Re}(E^{1})=\sum_{n=0}^{L_{x}} \left| \psi_{n}^{0} \right|^{2} \cos(2 \pi \alpha n)$ represents the real part of the correction to the eigenvalues, while $\operatorname{Im}(E^{1})=h \sum_{n=0}^{L_{x}} \left| \psi_{n}^{0} \right|^{2} \sin (2 \pi \alpha n)$ denotes the imaginary part. 

Notably, this correction also depends on the $\left| \psi_{n}^{0} \right|^{2}$ distribution, as seen in Eq. (\ref{perturbation energy}). Additionally, the distribution of topological edge states at the boundary is energy-dependent. As we shift from the energy gap to higher or lower energy bands, the distribution of topological edge states inside the material becomes increasingly significant. Consequently, the energy spectrum follows an elliptic function dependence that is governed by these edge states, expressed as:
\begin{align}
\frac{\left[\operatorname{Re}(E^{0}) + \operatorname{Re}(E^{1})\right]^2}{a^2}+\frac{\left[\operatorname{Im}(E^{1}) \right]^2}{b^2}=1.
\label{eq:half-ellipse}
\end{align}
The discrepancy between the theory and the numerical results is attributed to the higher order perturbation, as shown in Fig. \ref{fig: weak}(a). As the imaginary potential becomes smaller, the perturbation theory aligns more closely with the numerical outcomes

Due to the inclusion of gain/loss domain walls in the up boundary, the effective chiral Hamiltonian in Eq. (\ref{eq: low Hamiltonian}) is substituted with $H_{chiral} = v_{f}k +i V + \text{domain}$. And the Schr\"{o}dinger equation with OBC can be rewritten as
\begin{align}
\left[-i v_f \frac{d}{d x}+ \text{domain} +i V(x)\right] \psi(x)=(\epsilon_{r}+i \epsilon_{i})  \psi(x).
\end{align}
The solution of the wave function is given by
\begin{small}
\begin{align}
\psi(x)=\frac{1}{\sqrt{C}} \exp \left(i \frac{\epsilon_{r} }{v_f} x\right) \exp \left(\int_0^{L_{x}} d x^{\prime} \frac{ \text{domain} + V \left(x^{\prime}\right)-\epsilon_{i}}{v_f}\right).
\end{align}
\end{small}
Here $\epsilon_i$ is still the average imaginary potential, see Eq. (\ref{eq: epsilon_i}). Since the domain walls have opposite signs, $\int_{0}^{L_x} \text{domain}=0$.

To validate our prediction, we performed numerical calculations (blue line), which closely match our analytical results (red star-line), as shown in Fig. \ref{fig: weak}(b). Consequently, $H_{G} (H_{L})$ still exhibits localized topological edge states due to the inherent topological properties of the photonic topological insulator, in the presence of non-Hermitian effects. Additionally, the localization length of the wave function in Fig. \ref{fig: weak}(b) is inversely proportional to $\frac{1}{\xi} \propto \text{domain} + V \left(x^{\prime}\right)-\epsilon_{i}$.

\section{PHASE TRANSITION IN IMAGINARY POTENTIAL}
\label{PT phase transition}
\begin{figure}[]
	\centering
	\includegraphics[width=3.5in]{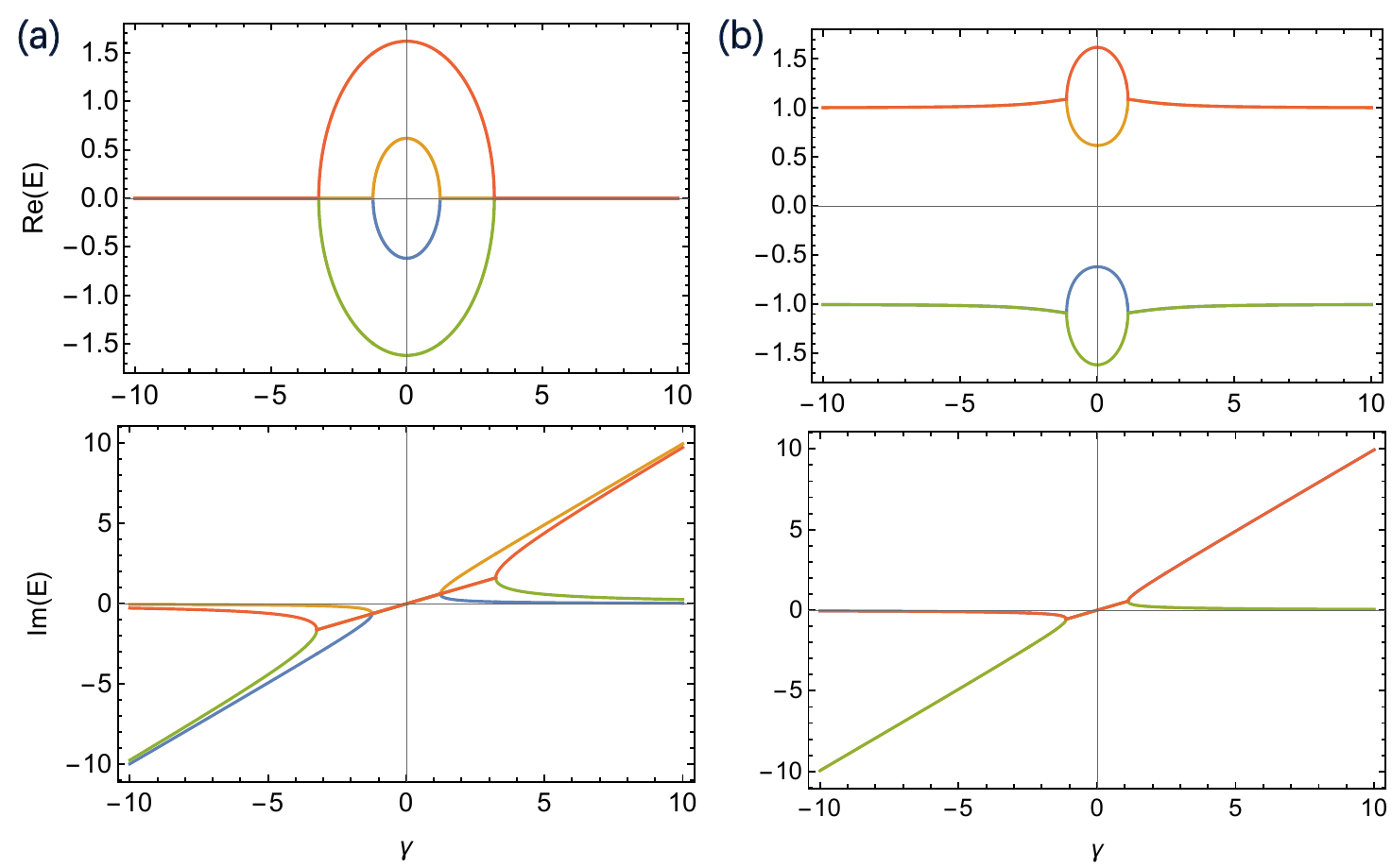}
	\caption{The energy spectrum of the simplest four lattice model. (a) The real and imaginary parts of the energy spectrum of $H_a$ as a function of $\gamma$. (b) The real and imaginary parts of the energy spectrum of $H_b$ as a function of $\gamma$.}
\label{fig: PT}
\end{figure}

We consider a four lattices model [Eq. (\ref{eq: HNN})] where two non-adjacent lattices are dissipative and the other two are not, and the Hamiltonian $H_{NNN}=H_{a}$
\begin{equation}
H_{a}=\left(
\begin{array}{cccc}
 i \gamma  & t & 0 & 0 \\
 t & 0 & t & 0 \\
 0 & t & i \gamma  & t \\
 0 & 0 & t & 0 \\
\end{array}
\right),
\end{equation}
where the $t$ is the nearest hopping term and the $\gamma$ is the non-adjacent on-site imaginary potential. $H_{a}$ is PT-symmetric, i.e., $PT H_{a}(P T)^{-1}=H_{a}$. The exceptional points are $\gamma=(\sqrt{5} \pm 1) t$, and the eigenvalues are
\begin{align}
& \lambda_1=\frac{1}{2} \left(-\sqrt{-\gamma ^2-2 \left(\sqrt{5}-3\right) t^2}+i \gamma \right), \notag\\
& \lambda_2=\frac{1}{2} \left(\sqrt{-\gamma ^2-2 \left(\sqrt{5}-3\right) t^2}+i \gamma \right), \notag\\
& \lambda_3=\frac{1}{2} \left(-\sqrt{2 \left(\sqrt{5}+3\right) t^2-\gamma ^2}+i \gamma \right),  \notag\\
& \lambda_4=\frac{1}{2} \left(\sqrt{2 \left(\sqrt{5}+3\right) t^2-\gamma ^2}+i \gamma \right).
\end{align}
As the eigenvalues are symmetric, we only need to consider $\lambda_i > 0$ ($i=1, 2, 3, 4$), see Fig.\ref{fig: PT} (a). The system hold $PT$ symmetry when $\gamma<(\sqrt{5}-1)t$, where $Im(\lambda_i)$ are all equal to each other. When $(\sqrt{5}-1)t<\gamma<(\sqrt{5}+1)t$, the imaginary part $Im(\lambda_1,2)$ are different, so $\gamma=(\sqrt{5}-1)t$ is the first critical point. Moreover, the $PT$-symmetry is broken again when $\gamma>(\sqrt{5} +1)t$, that the real part of the four modes vanishes.

Conversely, the Hamiltonian with two adjacent imaginary potentials is
\begin{equation}
H_{b}=\left(
\begin{array}{cccc}
 i \gamma  & t & 0 & 0 \\
 t &  i \gamma & t & 0 \\
 0 & t & 0  & t \\
 0 & 0 & t & 0 \\
\end{array}
\right),
\end{equation}
where $H_b$ is also PT-symmetric. Then the eigenvalues of $H_b$ is
\begin{align}
& \lambda_1=\frac{1}{2} \left(-\sqrt{-\gamma ^2-2 t \sqrt{5 t^2-4 \gamma ^2}+6 t^2}+i \gamma \right), \notag\\
& \lambda_2=\frac{1}{2} \left(\sqrt{-\gamma ^2-2 t \sqrt{5 t^2-4 \gamma ^2}+6 t^2}+i \gamma \right), \notag\\
& \lambda_3=\frac{1}{2} \left(-\sqrt{-\gamma ^2+2 t \sqrt{5 t^2-4 \gamma ^2}+6 t^2}+i \gamma \right),  \notag\\
& \lambda_4=\frac{1}{2} \left(\sqrt{-\gamma ^2+2 t \sqrt{5 t^2-4 \gamma ^2}+6 t^2}+i \gamma \right).
\end{align}
However, there is only one exceptional point where $Im(\lambda_1)=Im(\lambda_2)$ and $Im(\lambda_3)=Im(\lambda_4)$, as shown in Fig. \ref{fig: PT}(b).
\newpage
\nocite{*}

\bibliography{NH_chiral_quasipriodic}

\begin{thebibliography}{77}%
\makeatletter
\providecommand \@ifxundefined [1]{%
 \@ifx{#1\undefined}
}%
\providecommand \@ifnum [1]{%
 \ifnum #1\expandafter \@firstoftwo
 \else \expandafter \@secondoftwo
 \fi
}%
\providecommand \@ifx [1]{%
 \ifx #1\expandafter \@firstoftwo
 \else \expandafter \@secondoftwo
 \fi
}%
\providecommand \natexlab [1]{#1}%
\providecommand \enquote  [1]{``#1''}%
\providecommand \bibnamefont  [1]{#1}%
\providecommand \bibfnamefont [1]{#1}%
\providecommand \citenamefont [1]{#1}%
\providecommand \href@noop [0]{\@secondoftwo}%
\providecommand \href [0]{\begingroup \@sanitize@url \@href}%
\providecommand \@href[1]{\@@startlink{#1}\@@href}%
\providecommand \@@href[1]{\endgroup#1\@@endlink}%
\providecommand \@sanitize@url [0]{\catcode `\\12\catcode `\$12\catcode
  `\&12\catcode `\#12\catcode `\^12\catcode `\_12\catcode `\%12\relax}%
\providecommand \@@startlink[1]{}%
\providecommand \@@endlink[0]{}%
\providecommand \url  [0]{\begingroup\@sanitize@url \@url }%
\providecommand \@url [1]{\endgroup\@href {#1}{\urlprefix }}%
\providecommand \urlprefix  [0]{URL }%
\providecommand \Eprint [0]{\href }%
\providecommand \doibase [0]{https://doi.org/}%
\providecommand \selectlanguage [0]{\@gobble}%
\providecommand \bibinfo  [0]{\@secondoftwo}%
\providecommand \bibfield  [0]{\@secondoftwo}%
\providecommand \translation [1]{[#1]}%
\providecommand \BibitemOpen [0]{}%
\providecommand \bibitemStop [0]{}%
\providecommand \bibitemNoStop [0]{.\EOS\space}%
\providecommand \EOS [0]{\spacefactor3000\relax}%
\providecommand \BibitemShut  [1]{\csname bibitem#1\endcsname}%
\let\auto@bib@innerbib\@empty
\bibitem [{\citenamefont {Carmichael}(1993)}]{non-Hermitian1}%
  \BibitemOpen
  \bibfield  {author} {\bibinfo {author} {\bibfnamefont {H.~J.}\ \bibnamefont
  {Carmichael}},\ }\bibfield  {title} {\bibinfo {title} {Quantum trajectory
  theory for cascaded open systems},\ }\href
  {https://doi.org/10.1103/PhysRevLett.70.2273} {\bibfield  {journal} {\bibinfo
   {journal} {Phys. Rev. Lett.}\ }\textbf {\bibinfo {volume} {70}},\ \bibinfo
  {pages} {2273} (\bibinfo {year} {1993})}\BibitemShut {NoStop}%
\bibitem [{\citenamefont {Bender}(2007)}]{non-Hermitian2}%
  \BibitemOpen
  \bibfield  {author} {\bibinfo {author} {\bibfnamefont {C.~M.}\ \bibnamefont
  {Bender}},\ }\bibfield  {title} {\bibinfo {title} {Making sense of
  non-hermitian hamiltonians},\ }\href
  {https://doi.org/10.1088/0034-4885/70/6/R03} {\bibfield  {journal} {\bibinfo
  {journal} {Reports on Progress in Physics}\ }\textbf {\bibinfo {volume}
  {70}},\ \bibinfo {pages} {947} (\bibinfo {year} {2007})}\BibitemShut
  {NoStop}%
\bibitem [{\citenamefont {El-Ganainy}\ \emph {et~al.}(2018)\citenamefont
  {El-Ganainy}, \citenamefont {Makris}, \citenamefont {Khajavikhan},
  \citenamefont {Musslimani}, \citenamefont {Rotter},\ and\ \citenamefont
  {Christodoulides}}]{non-Hermitian3}%
  \BibitemOpen
  \bibfield  {author} {\bibinfo {author} {\bibfnamefont {R.}~\bibnamefont
  {El-Ganainy}}, \bibinfo {author} {\bibfnamefont {K.~G.}\ \bibnamefont
  {Makris}}, \bibinfo {author} {\bibfnamefont {M.}~\bibnamefont {Khajavikhan}},
  \bibinfo {author} {\bibfnamefont {Z.~H.}\ \bibnamefont {Musslimani}},
  \bibinfo {author} {\bibfnamefont {S.}~\bibnamefont {Rotter}},\ and\ \bibinfo
  {author} {\bibfnamefont {D.~N.}\ \bibnamefont {Christodoulides}},\ }\bibfield
   {title} {\bibinfo {title} {Non-hermitian physics and pt symmetry},\ }\href
  {https://doi.org/10.1038/nphys4323} {\bibfield  {journal} {\bibinfo
  {journal} {Nature Physics}\ }\textbf {\bibinfo {volume} {14}},\ \bibinfo
  {pages} {11} (\bibinfo {year} {2018})}\BibitemShut {NoStop}%
\bibitem [{\citenamefont {Bender}\ and\ \citenamefont
  {Boettcher}(1998)}]{non-Hermitian4}%
  \BibitemOpen
  \bibfield  {author} {\bibinfo {author} {\bibfnamefont {C.~M.}\ \bibnamefont
  {Bender}}\ and\ \bibinfo {author} {\bibfnamefont {S.}~\bibnamefont
  {Boettcher}},\ }\bibfield  {title} {\bibinfo {title} {Real spectra in
  non-hermitian hamiltonians having $\mathcal{P}\mathcal{T}$ symmetry},\ }\href
  {https://doi.org/10.1103/PhysRevLett.80.5243} {\bibfield  {journal} {\bibinfo
   {journal} {Phys. Rev. Lett.}\ }\textbf {\bibinfo {volume} {80}},\ \bibinfo
  {pages} {5243} (\bibinfo {year} {1998})}\BibitemShut {NoStop}%
\bibitem [{\citenamefont {Miri}\ and\ \citenamefont
  {Alù}(2019)}]{non-Hermitian5}%
  \BibitemOpen
  \bibfield  {author} {\bibinfo {author} {\bibfnamefont {M.-A.}\ \bibnamefont
  {Miri}}\ and\ \bibinfo {author} {\bibfnamefont {A.}~\bibnamefont {Alù}},\
  }\bibfield  {title} {\bibinfo {title} {Exceptional points in optics and
  photonics},\ }\href {https://doi.org/10.1126/science.aar7709} {\bibfield
  {journal} {\bibinfo  {journal} {Science}\ }\textbf {\bibinfo {volume}
  {363}},\ \bibinfo {pages} {eaar7709} (\bibinfo {year} {2019})}\BibitemShut
  {NoStop}%
\bibitem [{\citenamefont {Ashida}\ \emph
  {et~al.}(2020{\natexlab{a}})\citenamefont {Ashida}, \citenamefont {Gong},\
  and\ \citenamefont {Ueda}}]{non-Hermitian6}%
  \BibitemOpen
  \bibfield  {author} {\bibinfo {author} {\bibfnamefont {Y.}~\bibnamefont
  {Ashida}}, \bibinfo {author} {\bibfnamefont {Z.}~\bibnamefont {Gong}},\ and\
  \bibinfo {author} {\bibfnamefont {M.}~\bibnamefont {Ueda}},\ }\bibfield
  {title} {\bibinfo {title} {Non-hermitian physics},\ }\href
  {https://doi.org/10.1080/00018732.2021.1876991} {\bibfield  {journal}
  {\bibinfo  {journal} {Advances in Physics}\ }\textbf {\bibinfo {volume}
  {69}},\ \bibinfo {pages} {249} (\bibinfo {year}
  {2020}{\natexlab{a}})}\BibitemShut {NoStop}%
\bibitem [{\citenamefont {Yao}\ and\ \citenamefont
  {Wang}(2018{\natexlab{a}})}]{non-Hermitian7}%
  \BibitemOpen
  \bibfield  {author} {\bibinfo {author} {\bibfnamefont {S.}~\bibnamefont
  {Yao}}\ and\ \bibinfo {author} {\bibfnamefont {Z.}~\bibnamefont {Wang}},\
  }\bibfield  {title} {\bibinfo {title} {Edge states and topological invariants
  of non-hermitian systems},\ }\href
  {https://doi.org/10.1103/PhysRevLett.121.086803} {\bibfield  {journal}
  {\bibinfo  {journal} {Phys. Rev. Lett.}\ }\textbf {\bibinfo {volume} {121}},\
  \bibinfo {pages} {086803} (\bibinfo {year} {2018}{\natexlab{a}})}\BibitemShut
  {NoStop}%
\bibitem [{\citenamefont {Kawabata}\ \emph {et~al.}(2019)\citenamefont
  {Kawabata}, \citenamefont {Shiozaki}, \citenamefont {Ueda},\ and\
  \citenamefont {Sato}}]{non-Hermitian8}%
  \BibitemOpen
  \bibfield  {author} {\bibinfo {author} {\bibfnamefont {K.}~\bibnamefont
  {Kawabata}}, \bibinfo {author} {\bibfnamefont {K.}~\bibnamefont {Shiozaki}},
  \bibinfo {author} {\bibfnamefont {M.}~\bibnamefont {Ueda}},\ and\ \bibinfo
  {author} {\bibfnamefont {M.}~\bibnamefont {Sato}},\ }\bibfield  {title}
  {\bibinfo {title} {Symmetry and topology in non-hermitian physics},\ }\href
  {https://doi.org/10.1103/PhysRevX.9.041015} {\bibfield  {journal} {\bibinfo
  {journal} {Phys. Rev. X}\ }\textbf {\bibinfo {volume} {9}},\ \bibinfo {pages}
  {041015} (\bibinfo {year} {2019})}\BibitemShut {NoStop}%
\bibitem [{\citenamefont {Zhou}\ and\ \citenamefont
  {Lee}(2019)}]{non-Hermitian9}%
  \BibitemOpen
  \bibfield  {author} {\bibinfo {author} {\bibfnamefont {H.}~\bibnamefont
  {Zhou}}\ and\ \bibinfo {author} {\bibfnamefont {J.~Y.}\ \bibnamefont {Lee}},\
  }\bibfield  {title} {\bibinfo {title} {Periodic table for topological bands
  with non-hermitian symmetries},\ }\href
  {https://doi.org/10.1103/PhysRevB.99.235112} {\bibfield  {journal} {\bibinfo
  {journal} {Phys. Rev. B}\ }\textbf {\bibinfo {volume} {99}},\ \bibinfo
  {pages} {235112} (\bibinfo {year} {2019})}\BibitemShut {NoStop}%
\bibitem [{\citenamefont {Ghatak}\ and\ \citenamefont
  {Das}(2019)}]{non-Hermitian10}%
  \BibitemOpen
  \bibfield  {author} {\bibinfo {author} {\bibfnamefont {A.}~\bibnamefont
  {Ghatak}}\ and\ \bibinfo {author} {\bibfnamefont {T.}~\bibnamefont {Das}},\
  }\bibfield  {title} {\bibinfo {title} {New topological invariants in
  non-hermitian systems},\ }\href {https://doi.org/10.1088/1361-648X/ab11b3}
  {\bibfield  {journal} {\bibinfo  {journal} {Journal of Physics: Condensed
  Matter}\ }\textbf {\bibinfo {volume} {31}},\ \bibinfo {pages} {263001}
  (\bibinfo {year} {2019})}\BibitemShut {NoStop}%
\bibitem [{\citenamefont {Yao}\ and\ \citenamefont
  {Wang}(2018{\natexlab{b}})}]{non-Hermitian11}%
  \BibitemOpen
  \bibfield  {author} {\bibinfo {author} {\bibfnamefont {S.}~\bibnamefont
  {Yao}}\ and\ \bibinfo {author} {\bibfnamefont {Z.}~\bibnamefont {Wang}},\
  }\bibfield  {title} {\bibinfo {title} {Edge states and topological invariants
  of non-hermitian systems},\ }\href
  {https://doi.org/10.1103/PhysRevLett.121.086803} {\bibfield  {journal}
  {\bibinfo  {journal} {Phys. Rev. Lett.}\ }\textbf {\bibinfo {volume} {121}},\
  \bibinfo {pages} {086803} (\bibinfo {year} {2018}{\natexlab{b}})}\BibitemShut
  {NoStop}%
\bibitem [{\citenamefont {Yao}\ \emph {et~al.}(2018)\citenamefont {Yao},
  \citenamefont {Song},\ and\ \citenamefont {Wang}}]{non-Bloch_BBC1}%
  \BibitemOpen
  \bibfield  {author} {\bibinfo {author} {\bibfnamefont {S.}~\bibnamefont
  {Yao}}, \bibinfo {author} {\bibfnamefont {F.}~\bibnamefont {Song}},\ and\
  \bibinfo {author} {\bibfnamefont {Z.}~\bibnamefont {Wang}},\ }\bibfield
  {title} {\bibinfo {title} {Non-hermitian chern bands},\ }\href
  {https://doi.org/10.1103/PhysRevLett.121.136802} {\bibfield  {journal}
  {\bibinfo  {journal} {Phys. Rev. Lett.}\ }\textbf {\bibinfo {volume} {121}},\
  \bibinfo {pages} {136802} (\bibinfo {year} {2018})}\BibitemShut {NoStop}%
\bibitem [{\citenamefont {Kunitski}\ \emph {et~al.}(2019)\citenamefont
  {Kunitski}, \citenamefont {Eicke}, \citenamefont {Huber}, \citenamefont
  {K{\"o}hler}, \citenamefont {Zeller}, \citenamefont {Voigtsberger},
  \citenamefont {Schlott}, \citenamefont {Henrichs}, \citenamefont {Sann},
  \citenamefont {Trinter}, \citenamefont {Schmidt}, \citenamefont {Kalinin},
  \citenamefont {Sch{\"o}ffler}, \citenamefont {Jahnke}, \citenamefont {Lein},\
  and\ \citenamefont {D{\"o}rner}}]{non-Hermitian12}%
  \BibitemOpen
  \bibfield  {author} {\bibinfo {author} {\bibfnamefont {M.}~\bibnamefont
  {Kunitski}}, \bibinfo {author} {\bibfnamefont {N.}~\bibnamefont {Eicke}},
  \bibinfo {author} {\bibfnamefont {P.}~\bibnamefont {Huber}}, \bibinfo
  {author} {\bibfnamefont {J.}~\bibnamefont {K{\"o}hler}}, \bibinfo {author}
  {\bibfnamefont {S.}~\bibnamefont {Zeller}}, \bibinfo {author} {\bibfnamefont
  {J.}~\bibnamefont {Voigtsberger}}, \bibinfo {author} {\bibfnamefont
  {N.}~\bibnamefont {Schlott}}, \bibinfo {author} {\bibfnamefont
  {K.}~\bibnamefont {Henrichs}}, \bibinfo {author} {\bibfnamefont
  {H.}~\bibnamefont {Sann}}, \bibinfo {author} {\bibfnamefont {F.}~\bibnamefont
  {Trinter}}, \bibinfo {author} {\bibfnamefont {L.~P.~H.}\ \bibnamefont
  {Schmidt}}, \bibinfo {author} {\bibfnamefont {A.}~\bibnamefont {Kalinin}},
  \bibinfo {author} {\bibfnamefont {M.~S.}\ \bibnamefont {Sch{\"o}ffler}},
  \bibinfo {author} {\bibfnamefont {T.}~\bibnamefont {Jahnke}}, \bibinfo
  {author} {\bibfnamefont {M.}~\bibnamefont {Lein}},\ and\ \bibinfo {author}
  {\bibfnamefont {R.}~\bibnamefont {D{\"o}rner}},\ }\bibfield  {title}
  {\bibinfo {title} {Double-slit photoelectron interference in strong-field
  ionization of the neon dimer},\ }\href
  {https://doi.org/10.1038/s41467-018-07882-8} {\bibfield  {journal} {\bibinfo
  {journal} {Nature Communications}\ }\textbf {\bibinfo {volume} {10}},\
  \bibinfo {pages} {1} (\bibinfo {year} {2019})}\BibitemShut {NoStop}%
\bibitem [{\citenamefont {Yokomizo}\ and\ \citenamefont
  {Murakami}(2019)}]{non-Hermitian_skin_effects1}%
  \BibitemOpen
  \bibfield  {author} {\bibinfo {author} {\bibfnamefont {K.}~\bibnamefont
  {Yokomizo}}\ and\ \bibinfo {author} {\bibfnamefont {S.}~\bibnamefont
  {Murakami}},\ }\bibfield  {title} {\bibinfo {title} {Non-bloch band theory of
  non-hermitian systems},\ }\href
  {https://doi.org/10.1103/PhysRevLett.123.066404} {\bibfield  {journal}
  {\bibinfo  {journal} {Phys. Rev. Lett.}\ }\textbf {\bibinfo {volume} {123}},\
  \bibinfo {pages} {066404} (\bibinfo {year} {2019})}\BibitemShut {NoStop}%
\bibitem [{\citenamefont {Lee}\ and\ \citenamefont
  {Thomale}(2019)}]{non-Hermitian_skin_effects2}%
  \BibitemOpen
  \bibfield  {author} {\bibinfo {author} {\bibfnamefont {C.~H.}\ \bibnamefont
  {Lee}}\ and\ \bibinfo {author} {\bibfnamefont {R.}~\bibnamefont {Thomale}},\
  }\bibfield  {title} {\bibinfo {title} {Anatomy of skin modes and topology in
  non-hermitian systems},\ }\href {https://doi.org/10.1103/PhysRevB.99.201103}
  {\bibfield  {journal} {\bibinfo  {journal} {Phys. Rev. B}\ }\textbf {\bibinfo
  {volume} {99}},\ \bibinfo {pages} {201103} (\bibinfo {year}
  {2019})}\BibitemShut {NoStop}%
\bibitem [{\citenamefont {Kunst}\ \emph {et~al.}(2018)\citenamefont {Kunst},
  \citenamefont {Edvardsson}, \citenamefont {Budich},\ and\ \citenamefont
  {Bergholtz}}]{non-Hermitian_skin_effects3}%
  \BibitemOpen
  \bibfield  {author} {\bibinfo {author} {\bibfnamefont {F.~K.}\ \bibnamefont
  {Kunst}}, \bibinfo {author} {\bibfnamefont {E.}~\bibnamefont {Edvardsson}},
  \bibinfo {author} {\bibfnamefont {J.~C.}\ \bibnamefont {Budich}},\ and\
  \bibinfo {author} {\bibfnamefont {E.~J.}\ \bibnamefont {Bergholtz}},\
  }\bibfield  {title} {\bibinfo {title} {Biorthogonal bulk-boundary
  correspondence in non-hermitian systems},\ }\href
  {https://doi.org/10.1103/PhysRevLett.121.026808} {\bibfield  {journal}
  {\bibinfo  {journal} {Phys. Rev. Lett.}\ }\textbf {\bibinfo {volume} {121}},\
  \bibinfo {pages} {026808} (\bibinfo {year} {2018})}\BibitemShut {NoStop}%
\bibitem [{\citenamefont {Borgnia}\ \emph {et~al.}(2020)\citenamefont
  {Borgnia}, \citenamefont {Kruchkov},\ and\ \citenamefont
  {Slager}}]{non-Hermitian_skin_effects4}%
  \BibitemOpen
  \bibfield  {author} {\bibinfo {author} {\bibfnamefont {D.~S.}\ \bibnamefont
  {Borgnia}}, \bibinfo {author} {\bibfnamefont {A.~J.}\ \bibnamefont
  {Kruchkov}},\ and\ \bibinfo {author} {\bibfnamefont {R.-J.}\ \bibnamefont
  {Slager}},\ }\bibfield  {title} {\bibinfo {title} {Non-hermitian boundary
  modes and topology},\ }\href {https://doi.org/10.1103/PhysRevLett.124.056802}
  {\bibfield  {journal} {\bibinfo  {journal} {Phys. Rev. Lett.}\ }\textbf
  {\bibinfo {volume} {124}},\ \bibinfo {pages} {056802} (\bibinfo {year}
  {2020})}\BibitemShut {NoStop}%
\bibitem [{\citenamefont {McDonald}\ \emph {et~al.}(2018)\citenamefont
  {McDonald}, \citenamefont {Pereg-Barnea},\ and\ \citenamefont
  {Clerk}}]{non-Hermitian_skin_effects5}%
  \BibitemOpen
  \bibfield  {author} {\bibinfo {author} {\bibfnamefont {A.}~\bibnamefont
  {McDonald}}, \bibinfo {author} {\bibfnamefont {T.}~\bibnamefont
  {Pereg-Barnea}},\ and\ \bibinfo {author} {\bibfnamefont {A.~A.}\ \bibnamefont
  {Clerk}},\ }\bibfield  {title} {\bibinfo {title} {Phase-dependent chiral
  transport and effective non-hermitian dynamics in a bosonic kitaev-majorana
  chain},\ }\href {https://doi.org/10.1103/PhysRevX.8.041031} {\bibfield
  {journal} {\bibinfo  {journal} {Phys. Rev. X}\ }\textbf {\bibinfo {volume}
  {8}},\ \bibinfo {pages} {041031} (\bibinfo {year} {2018})}\BibitemShut
  {NoStop}%
\bibitem [{\citenamefont {Martinez~Alvarez}\ \emph {et~al.}(2018)\citenamefont
  {Martinez~Alvarez}, \citenamefont {Barrios~Vargas},\ and\ \citenamefont
  {Foa~Torres}}]{non-Hermitian_skin_effects6}%
  \BibitemOpen
  \bibfield  {author} {\bibinfo {author} {\bibfnamefont {V.~M.}\ \bibnamefont
  {Martinez~Alvarez}}, \bibinfo {author} {\bibfnamefont {J.~E.}\ \bibnamefont
  {Barrios~Vargas}},\ and\ \bibinfo {author} {\bibfnamefont {L.~E.~F.}\
  \bibnamefont {Foa~Torres}},\ }\bibfield  {title} {\bibinfo {title}
  {Non-hermitian robust edge states in one dimension: Anomalous localization
  and eigenspace condensation at exceptional points},\ }\href
  {https://doi.org/10.1103/PhysRevB.97.121401} {\bibfield  {journal} {\bibinfo
  {journal} {Phys. Rev. B}\ }\textbf {\bibinfo {volume} {97}},\ \bibinfo
  {pages} {121401} (\bibinfo {year} {2018})}\BibitemShut {NoStop}%
\bibitem [{\citenamefont {Zhang}\ \emph {et~al.}(2020)\citenamefont {Zhang},
  \citenamefont {Yang},\ and\ \citenamefont
  {Fang}}]{non-Hermitian_skin_effects7}%
  \BibitemOpen
  \bibfield  {author} {\bibinfo {author} {\bibfnamefont {K.}~\bibnamefont
  {Zhang}}, \bibinfo {author} {\bibfnamefont {Z.}~\bibnamefont {Yang}},\ and\
  \bibinfo {author} {\bibfnamefont {C.}~\bibnamefont {Fang}},\ }\bibfield
  {title} {\bibinfo {title} {Correspondence between winding numbers and skin
  modes in non-hermitian systems},\ }\href
  {https://doi.org/10.1103/PhysRevLett.125.126402} {\bibfield  {journal}
  {\bibinfo  {journal} {Phys. Rev. Lett.}\ }\textbf {\bibinfo {volume} {125}},\
  \bibinfo {pages} {126402} (\bibinfo {year} {2020})}\BibitemShut {NoStop}%
\bibitem [{\citenamefont
  {Longhi}(2019{\natexlab{a}})}]{non-Hermitian_skin_effects8}%
  \BibitemOpen
  \bibfield  {author} {\bibinfo {author} {\bibfnamefont {S.}~\bibnamefont
  {Longhi}},\ }\bibfield  {title} {\bibinfo {title} {Probing non-hermitian skin
  effect and non-bloch phase transitions},\ }\href
  {https://doi.org/10.1103/PhysRevResearch.1.023013} {\bibfield  {journal}
  {\bibinfo  {journal} {Phys. Rev. Res.}\ }\textbf {\bibinfo {volume} {1}},\
  \bibinfo {pages} {023013} (\bibinfo {year} {2019}{\natexlab{a}})}\BibitemShut
  {NoStop}%
\bibitem [{\citenamefont {Li}\ \emph {et~al.}(2020)\citenamefont {Li},
  \citenamefont {Lee}, \citenamefont {Mu},\ and\ \citenamefont
  {Gong}}]{non-Hermitian_skin_effects9}%
  \BibitemOpen
  \bibfield  {author} {\bibinfo {author} {\bibfnamefont {L.}~\bibnamefont
  {Li}}, \bibinfo {author} {\bibfnamefont {C.~H.}\ \bibnamefont {Lee}},
  \bibinfo {author} {\bibfnamefont {S.}~\bibnamefont {Mu}},\ and\ \bibinfo
  {author} {\bibfnamefont {J.}~\bibnamefont {Gong}},\ }\bibfield  {title}
  {\bibinfo {title} {Critical non-hermitian skin effect},\ }\href
  {https://doi.org/10.1038/s41467-020-18917-4} {\bibfield  {journal} {\bibinfo
  {journal} {Nature Communications}\ }\textbf {\bibinfo {volume} {11}},\
  \bibinfo {pages} {5491} (\bibinfo {year} {2020})}\BibitemShut {NoStop}%
\bibitem [{\citenamefont {Song}\ \emph {et~al.}(2019)\citenamefont {Song},
  \citenamefont {Sun}, \citenamefont {Chen}, \citenamefont {Song},
  \citenamefont {Xiao}, \citenamefont {Zhu},\ and\ \citenamefont
  {Li}}]{experiment1}%
  \BibitemOpen
  \bibfield  {author} {\bibinfo {author} {\bibfnamefont {W.}~\bibnamefont
  {Song}}, \bibinfo {author} {\bibfnamefont {W.}~\bibnamefont {Sun}}, \bibinfo
  {author} {\bibfnamefont {C.}~\bibnamefont {Chen}}, \bibinfo {author}
  {\bibfnamefont {Q.}~\bibnamefont {Song}}, \bibinfo {author} {\bibfnamefont
  {S.}~\bibnamefont {Xiao}}, \bibinfo {author} {\bibfnamefont {S.}~\bibnamefont
  {Zhu}},\ and\ \bibinfo {author} {\bibfnamefont {T.}~\bibnamefont {Li}},\
  }\bibfield  {title} {\bibinfo {title} {Breakup and recovery of topological
  zero modes in finite non-hermitian optical lattices},\ }\href
  {https://doi.org/10.1103/PhysRevLett.123.165701} {\bibfield  {journal}
  {\bibinfo  {journal} {Phys. Rev. Lett.}\ }\textbf {\bibinfo {volume} {123}},\
  \bibinfo {pages} {165701} (\bibinfo {year} {2019})}\BibitemShut {NoStop}%
\bibitem [{\citenamefont {Zhao}\ \emph {et~al.}(2019)\citenamefont {Zhao},
  \citenamefont {Qiao}, \citenamefont {Wu}, \citenamefont {Midya},
  \citenamefont {Longhi},\ and\ \citenamefont {Feng}}]{experiment2}%
  \BibitemOpen
  \bibfield  {author} {\bibinfo {author} {\bibfnamefont {H.}~\bibnamefont
  {Zhao}}, \bibinfo {author} {\bibfnamefont {X.}~\bibnamefont {Qiao}}, \bibinfo
  {author} {\bibfnamefont {T.}~\bibnamefont {Wu}}, \bibinfo {author}
  {\bibfnamefont {B.}~\bibnamefont {Midya}}, \bibinfo {author} {\bibfnamefont
  {S.}~\bibnamefont {Longhi}},\ and\ \bibinfo {author} {\bibfnamefont
  {L.}~\bibnamefont {Feng}},\ }\bibfield  {title} {\bibinfo {title}
  {Non-hermitian topological light steering},\ }\href
  {https://doi.org/10.1126/science.aay1064} {\bibfield  {journal} {\bibinfo
  {journal} {Science}\ }\textbf {\bibinfo {volume} {365}},\ \bibinfo {pages}
  {1163} (\bibinfo {year} {2019})}\BibitemShut {NoStop}%
\bibitem [{\citenamefont {Weidemann}\ \emph {et~al.}(2020)\citenamefont
  {Weidemann}, \citenamefont {Kremer}, \citenamefont {Helbig}, \citenamefont
  {Hofmann}, \citenamefont {Stegmaier}, \citenamefont {Greiter}, \citenamefont
  {Thomale},\ and\ \citenamefont {Szameit}}]{experiment3}%
  \BibitemOpen
  \bibfield  {author} {\bibinfo {author} {\bibfnamefont {S.}~\bibnamefont
  {Weidemann}}, \bibinfo {author} {\bibfnamefont {M.}~\bibnamefont {Kremer}},
  \bibinfo {author} {\bibfnamefont {T.}~\bibnamefont {Helbig}}, \bibinfo
  {author} {\bibfnamefont {T.}~\bibnamefont {Hofmann}}, \bibinfo {author}
  {\bibfnamefont {A.}~\bibnamefont {Stegmaier}}, \bibinfo {author}
  {\bibfnamefont {M.}~\bibnamefont {Greiter}}, \bibinfo {author} {\bibfnamefont
  {R.}~\bibnamefont {Thomale}},\ and\ \bibinfo {author} {\bibfnamefont
  {A.}~\bibnamefont {Szameit}},\ }\bibfield  {title} {\bibinfo {title}
  {Topological funneling of light},\ }\href
  {https://doi.org/10.1126/science.aaz8727} {\bibfield  {journal} {\bibinfo
  {journal} {Science}\ }\textbf {\bibinfo {volume} {368}},\ \bibinfo {pages}
  {311} (\bibinfo {year} {2020})}\BibitemShut {NoStop}%
\bibitem [{\citenamefont {Hu}\ \emph {et~al.}(2021)\citenamefont {Hu},
  \citenamefont {Zhang}, \citenamefont {Zhang}, \citenamefont {Zheng},
  \citenamefont {Xiong}, \citenamefont {Yue}, \citenamefont {Wang},
  \citenamefont {Xu}, \citenamefont {Cheng}, \citenamefont {Liu},\ and\
  \citenamefont {Christensen}}]{experiment4}%
  \BibitemOpen
  \bibfield  {author} {\bibinfo {author} {\bibfnamefont {B.}~\bibnamefont
  {Hu}}, \bibinfo {author} {\bibfnamefont {Z.}~\bibnamefont {Zhang}}, \bibinfo
  {author} {\bibfnamefont {H.}~\bibnamefont {Zhang}}, \bibinfo {author}
  {\bibfnamefont {L.}~\bibnamefont {Zheng}}, \bibinfo {author} {\bibfnamefont
  {W.}~\bibnamefont {Xiong}}, \bibinfo {author} {\bibfnamefont
  {Z.}~\bibnamefont {Yue}}, \bibinfo {author} {\bibfnamefont {X.}~\bibnamefont
  {Wang}}, \bibinfo {author} {\bibfnamefont {J.}~\bibnamefont {Xu}}, \bibinfo
  {author} {\bibfnamefont {Y.}~\bibnamefont {Cheng}}, \bibinfo {author}
  {\bibfnamefont {X.}~\bibnamefont {Liu}},\ and\ \bibinfo {author}
  {\bibfnamefont {J.}~\bibnamefont {Christensen}},\ }\bibfield  {title}
  {\bibinfo {title} {Non-hermitian topological whispering gallery},\ }\href
  {https://doi.org/10.1038/s41586-021-03833-4} {\bibfield  {journal} {\bibinfo
  {journal} {Nature}\ }\textbf {\bibinfo {volume} {597}},\ \bibinfo {pages}
  {655} (\bibinfo {year} {2021})}\BibitemShut {NoStop}%
\bibitem [{\citenamefont {Öztürk}\ \emph {et~al.}(2021)\citenamefont
  {Öztürk}, \citenamefont {Lappe}, \citenamefont {Hellmann}, \citenamefont
  {Schmitt}, \citenamefont {Klaers}, \citenamefont {Vewinger}, \citenamefont
  {Kroha},\ and\ \citenamefont {Weitz}}]{experiment5}%
  \BibitemOpen
  \bibfield  {author} {\bibinfo {author} {\bibfnamefont {F.~E.}\ \bibnamefont
  {Öztürk}}, \bibinfo {author} {\bibfnamefont {T.}~\bibnamefont {Lappe}},
  \bibinfo {author} {\bibfnamefont {G.}~\bibnamefont {Hellmann}}, \bibinfo
  {author} {\bibfnamefont {J.}~\bibnamefont {Schmitt}}, \bibinfo {author}
  {\bibfnamefont {J.}~\bibnamefont {Klaers}}, \bibinfo {author} {\bibfnamefont
  {F.}~\bibnamefont {Vewinger}}, \bibinfo {author} {\bibfnamefont
  {J.}~\bibnamefont {Kroha}},\ and\ \bibinfo {author} {\bibfnamefont
  {M.}~\bibnamefont {Weitz}},\ }\bibfield  {title} {\bibinfo {title}
  {Observation of a non-hermitian phase transition in an optical quantum gas},\
  }\href {https://doi.org/10.1126/science.abe9869} {\bibfield  {journal}
  {\bibinfo  {journal} {Science}\ }\textbf {\bibinfo {volume} {372}},\ \bibinfo
  {pages} {88} (\bibinfo {year} {2021})}\BibitemShut {NoStop}%
\bibitem [{\citenamefont {Wang}\ \emph
  {et~al.}(2021{\natexlab{a}})\citenamefont {Wang}, \citenamefont {Dutt},
  \citenamefont {Yang}, \citenamefont {Wojcik}, \citenamefont {Vučković},\
  and\ \citenamefont {Fan}}]{experiment6}%
  \BibitemOpen
  \bibfield  {author} {\bibinfo {author} {\bibfnamefont {K.}~\bibnamefont
  {Wang}}, \bibinfo {author} {\bibfnamefont {A.}~\bibnamefont {Dutt}}, \bibinfo
  {author} {\bibfnamefont {K.~Y.}\ \bibnamefont {Yang}}, \bibinfo {author}
  {\bibfnamefont {C.~C.}\ \bibnamefont {Wojcik}}, \bibinfo {author}
  {\bibfnamefont {J.}~\bibnamefont {Vučković}},\ and\ \bibinfo {author}
  {\bibfnamefont {S.}~\bibnamefont {Fan}},\ }\bibfield  {title} {\bibinfo
  {title} {Generating arbitrary topological windings of a non-hermitian band},\
  }\href {https://doi.org/10.1126/science.abf6568} {\bibfield  {journal}
  {\bibinfo  {journal} {Science}\ }\textbf {\bibinfo {volume} {371}},\ \bibinfo
  {pages} {1240} (\bibinfo {year} {2021}{\natexlab{a}})}\BibitemShut {NoStop}%
\bibitem [{\citenamefont {Xiao}\ \emph
  {et~al.}(2021{\natexlab{a}})\citenamefont {Xiao}, \citenamefont {Deng},
  \citenamefont {Wang}, \citenamefont {Wang}, \citenamefont {Yi},\ and\
  \citenamefont {Xue}}]{experiment7}%
  \BibitemOpen
  \bibfield  {author} {\bibinfo {author} {\bibfnamefont {L.}~\bibnamefont
  {Xiao}}, \bibinfo {author} {\bibfnamefont {T.}~\bibnamefont {Deng}}, \bibinfo
  {author} {\bibfnamefont {K.}~\bibnamefont {Wang}}, \bibinfo {author}
  {\bibfnamefont {Z.}~\bibnamefont {Wang}}, \bibinfo {author} {\bibfnamefont
  {W.}~\bibnamefont {Yi}},\ and\ \bibinfo {author} {\bibfnamefont
  {P.}~\bibnamefont {Xue}},\ }\bibfield  {title} {\bibinfo {title} {Observation
  of non-bloch parity-time symmetry and exceptional points},\ }\href
  {https://doi.org/10.1103/PhysRevLett.126.230402} {\bibfield  {journal}
  {\bibinfo  {journal} {Phys. Rev. Lett.}\ }\textbf {\bibinfo {volume} {126}},\
  \bibinfo {pages} {230402} (\bibinfo {year} {2021}{\natexlab{a}})}\BibitemShut
  {NoStop}%
\bibitem [{\citenamefont {Lin}\ \emph {et~al.}(2022)\citenamefont {Lin},
  \citenamefont {Li}, \citenamefont {Xiao}, \citenamefont {Wang}, \citenamefont
  {Yi},\ and\ \citenamefont {Xue}}]{experiment8}%
  \BibitemOpen
  \bibfield  {author} {\bibinfo {author} {\bibfnamefont {Q.}~\bibnamefont
  {Lin}}, \bibinfo {author} {\bibfnamefont {T.}~\bibnamefont {Li}}, \bibinfo
  {author} {\bibfnamefont {L.}~\bibnamefont {Xiao}}, \bibinfo {author}
  {\bibfnamefont {K.}~\bibnamefont {Wang}}, \bibinfo {author} {\bibfnamefont
  {W.}~\bibnamefont {Yi}},\ and\ \bibinfo {author} {\bibfnamefont
  {P.}~\bibnamefont {Xue}},\ }\bibfield  {title} {\bibinfo {title} {Topological
  phase transitions and mobility edges in non-hermitian quasicrystals},\ }\href
  {https://doi.org/10.1103/PhysRevLett.129.113601} {\bibfield  {journal}
  {\bibinfo  {journal} {Phys. Rev. Lett.}\ }\textbf {\bibinfo {volume} {129}},\
  \bibinfo {pages} {113601} (\bibinfo {year} {2022})}\BibitemShut {NoStop}%
\bibitem [{\citenamefont {Lee}\ \emph {et~al.}(2019)\citenamefont {Lee},
  \citenamefont {Li},\ and\ \citenamefont
  {Gong}}]{hybrid_skin-topological_effect4}%
  \BibitemOpen
  \bibfield  {author} {\bibinfo {author} {\bibfnamefont {C.~H.}\ \bibnamefont
  {Lee}}, \bibinfo {author} {\bibfnamefont {L.}~\bibnamefont {Li}},\ and\
  \bibinfo {author} {\bibfnamefont {J.}~\bibnamefont {Gong}},\ }\bibfield
  {title} {\bibinfo {title} {Hybrid higher-order skin-topological modes in
  nonreciprocal systems},\ }\href
  {https://doi.org/10.1103/PhysRevLett.123.016805} {\bibfield  {journal}
  {\bibinfo  {journal} {Phys. Rev. Lett.}\ }\textbf {\bibinfo {volume} {123}},\
  \bibinfo {pages} {016805} (\bibinfo {year} {2019})}\BibitemShut {NoStop}%
\bibitem [{\citenamefont {Kawabata}\ \emph {et~al.}(2020)\citenamefont
  {Kawabata}, \citenamefont {Sato},\ and\ \citenamefont
  {Shiozaki}}]{hybrid_skin-topological_effect5}%
  \BibitemOpen
  \bibfield  {author} {\bibinfo {author} {\bibfnamefont {K.}~\bibnamefont
  {Kawabata}}, \bibinfo {author} {\bibfnamefont {M.}~\bibnamefont {Sato}},\
  and\ \bibinfo {author} {\bibfnamefont {K.}~\bibnamefont {Shiozaki}},\
  }\bibfield  {title} {\bibinfo {title} {Higher-order non-hermitian skin
  effect},\ }\href {https://doi.org/10.1103/PhysRevB.102.205118} {\bibfield
  {journal} {\bibinfo  {journal} {Phys. Rev. B}\ }\textbf {\bibinfo {volume}
  {102}},\ \bibinfo {pages} {205118} (\bibinfo {year} {2020})}\BibitemShut
  {NoStop}%
\bibitem [{\citenamefont {Li}\ \emph {et~al.}(2022)\citenamefont {Li},
  \citenamefont {Liang}, \citenamefont {Wang}, \citenamefont {Lu},\ and\
  \citenamefont {Liu}}]{hybrid_skin-topological_effect6}%
  \BibitemOpen
  \bibfield  {author} {\bibinfo {author} {\bibfnamefont {Y.}~\bibnamefont
  {Li}}, \bibinfo {author} {\bibfnamefont {C.}~\bibnamefont {Liang}}, \bibinfo
  {author} {\bibfnamefont {C.}~\bibnamefont {Wang}}, \bibinfo {author}
  {\bibfnamefont {C.}~\bibnamefont {Lu}},\ and\ \bibinfo {author}
  {\bibfnamefont {Y.-C.}\ \bibnamefont {Liu}},\ }\bibfield  {title} {\bibinfo
  {title} {Gain-loss-induced hybrid skin-topological effect},\ }\href
  {https://doi.org/10.1103/PhysRevLett.128.223903} {\bibfield  {journal}
  {\bibinfo  {journal} {Phys. Rev. Lett.}\ }\textbf {\bibinfo {volume} {128}},\
  \bibinfo {pages} {223903} (\bibinfo {year} {2022})}\BibitemShut {NoStop}%
\bibitem [{\citenamefont {Ma}\ \emph {et~al.}(2023)\citenamefont {Ma},
  \citenamefont {Cao}, \citenamefont {Wang}, \citenamefont {Wei},\ and\
  \citenamefont {Kou}}]{hybrid_skin-topological_effect1}%
  \BibitemOpen
  \bibfield  {author} {\bibinfo {author} {\bibfnamefont {X.}~\bibnamefont
  {Ma}}, \bibinfo {author} {\bibfnamefont {K.}~\bibnamefont {Cao}}, \bibinfo
  {author} {\bibfnamefont {X.}~\bibnamefont {Wang}}, \bibinfo {author}
  {\bibfnamefont {Z.}~\bibnamefont {Wei}},\ and\ \bibinfo {author}
  {\bibfnamefont {S.}~\bibnamefont {Kou}},\ }\href@noop {} {\bibinfo {title}
  {Non-hermitian chiral skin effect}} (\bibinfo {year} {2023}),\ \Eprint
  {https://arxiv.org/abs/2304.01422} {arXiv:2304.01422 [quant-ph]} \BibitemShut
  {NoStop}%
\bibitem [{\citenamefont {Yang}\ \emph {et~al.}(2023)\citenamefont {Yang},
  \citenamefont {Ren},\ and\ \citenamefont {peng
  Kou}}]{hybrid_skin-topological_effect2}%
  \BibitemOpen
  \bibfield  {author} {\bibinfo {author} {\bibfnamefont {F.}~\bibnamefont
  {Yang}}, \bibinfo {author} {\bibfnamefont {X.-P.}\ \bibnamefont {Ren}},\ and\
  \bibinfo {author} {\bibfnamefont {S.}~\bibnamefont {peng Kou}},\ }\href@noop
  {} {\bibinfo {title} {Non-hermitian chiral edge modes with complex fermi
  velocity}} (\bibinfo {year} {2023}),\ \Eprint
  {https://arxiv.org/abs/2307.14144} {arXiv:2307.14144 [cond-mat.str-el]}
  \BibitemShut {NoStop}%
\bibitem [{\citenamefont {Du}\ \emph {et~al.}(2023)\citenamefont {Du},
  \citenamefont {Ma},\ and\ \citenamefont
  {Kou}}]{hybrid_skin-topological_effect3}%
  \BibitemOpen
  \bibfield  {author} {\bibinfo {author} {\bibfnamefont {Q.}~\bibnamefont
  {Du}}, \bibinfo {author} {\bibfnamefont {X.-R.}\ \bibnamefont {Ma}},\ and\
  \bibinfo {author} {\bibfnamefont {S.-P.}\ \bibnamefont {Kou}},\ }\href@noop
  {} {\bibinfo {title} {Non-hermitian tearing by dissipation}} (\bibinfo {year}
  {2023}),\ \Eprint {https://arxiv.org/abs/2307.14340} {arXiv:2307.14340
  [cond-mat.mes-hall]} \BibitemShut {NoStop}%
\bibitem [{\citenamefont {Harper}(1955{\natexlab{a}})}]{AAH2}%
  \BibitemOpen
  \bibfield  {author} {\bibinfo {author} {\bibfnamefont {P.~G.}\ \bibnamefont
  {Harper}},\ }\bibfield  {title} {\bibinfo {title} {Single band motion of
  conduction electrons in a uniform magnetic field},\ }\href
  {https://doi.org/10.1088/0370-1298/68/10/304} {\bibfield  {journal} {\bibinfo
   {journal} {Proceedings of the Physical Society. Section A}\ }\textbf
  {\bibinfo {volume} {68}},\ \bibinfo {pages} {874} (\bibinfo {year}
  {1955}{\natexlab{a}})}\BibitemShut {NoStop}%
\bibitem [{\citenamefont {Aubry}\ and\ \citenamefont {Andr{\'e}}(1980)}]{AAH3}%
  \BibitemOpen
  \bibfield  {author} {\bibinfo {author} {\bibfnamefont {S.}~\bibnamefont
  {Aubry}}\ and\ \bibinfo {author} {\bibfnamefont {G.}~\bibnamefont
  {Andr{\'e}}},\ }\bibfield  {title} {\bibinfo {title} {Analyticity breaking
  and anderson localization in incommensurate lattices},\ }\href@noop {} {\
  (\bibinfo {year} {1980})}\BibitemShut {NoStop}%
\bibitem [{\citenamefont {Jotzu}\ \emph {et~al.}(2014)\citenamefont {Jotzu},
  \citenamefont {Messer}, \citenamefont {Desbuquois}, \citenamefont {Lebrat},
  \citenamefont {Uehlinger}, \citenamefont {Greif},\ and\ \citenamefont
  {Esslinger}}]{AAH4}%
  \BibitemOpen
  \bibfield  {author} {\bibinfo {author} {\bibfnamefont {G.}~\bibnamefont
  {Jotzu}}, \bibinfo {author} {\bibfnamefont {M.}~\bibnamefont {Messer}},
  \bibinfo {author} {\bibfnamefont {R.}~\bibnamefont {Desbuquois}}, \bibinfo
  {author} {\bibfnamefont {M.}~\bibnamefont {Lebrat}}, \bibinfo {author}
  {\bibfnamefont {T.}~\bibnamefont {Uehlinger}}, \bibinfo {author}
  {\bibfnamefont {D.}~\bibnamefont {Greif}},\ and\ \bibinfo {author}
  {\bibfnamefont {T.}~\bibnamefont {Esslinger}},\ }\bibfield  {title} {\bibinfo
  {title} {Experimental realization of the topological haldane model with
  ultracold fermions},\ }\href {https://doi.org/10.1038/nature13915} {\bibfield
   {journal} {\bibinfo  {journal} {Nature}\ }\textbf {\bibinfo {volume}
  {515}},\ \bibinfo {pages} {237} (\bibinfo {year} {2014})}\BibitemShut
  {NoStop}%
\bibitem [{\citenamefont {Roati}\ \emph {et~al.}(2008)\citenamefont {Roati},
  \citenamefont {D'Errico}, \citenamefont {Fallani}, \citenamefont {Fattori},
  \citenamefont {Fort}, \citenamefont {Zaccanti}, \citenamefont {Modugno},
  \citenamefont {Modugno},\ and\ \citenamefont {Inguscio}}]{AAH5}%
  \BibitemOpen
  \bibfield  {author} {\bibinfo {author} {\bibfnamefont {G.}~\bibnamefont
  {Roati}}, \bibinfo {author} {\bibfnamefont {C.}~\bibnamefont {D'Errico}},
  \bibinfo {author} {\bibfnamefont {L.}~\bibnamefont {Fallani}}, \bibinfo
  {author} {\bibfnamefont {M.}~\bibnamefont {Fattori}}, \bibinfo {author}
  {\bibfnamefont {C.}~\bibnamefont {Fort}}, \bibinfo {author} {\bibfnamefont
  {M.}~\bibnamefont {Zaccanti}}, \bibinfo {author} {\bibfnamefont
  {G.}~\bibnamefont {Modugno}}, \bibinfo {author} {\bibfnamefont
  {M.}~\bibnamefont {Modugno}},\ and\ \bibinfo {author} {\bibfnamefont
  {M.}~\bibnamefont {Inguscio}},\ }\bibfield  {title} {\bibinfo {title}
  {Anderson localization of a non-interacting bose--einstein condensate},\
  }\href {https://doi.org/10.1038/nature07071} {\bibfield  {journal} {\bibinfo
  {journal} {Nature}\ }\textbf {\bibinfo {volume} {453}},\ \bibinfo {pages}
  {895} (\bibinfo {year} {2008})}\BibitemShut {NoStop}%
\bibitem [{\citenamefont {Lahini}\ \emph {et~al.}(2009)\citenamefont {Lahini},
  \citenamefont {Pugatch}, \citenamefont {Pozzi}, \citenamefont {Sorel},
  \citenamefont {Morandotti}, \citenamefont {Davidson},\ and\ \citenamefont
  {Silberberg}}]{AAH6}%
  \BibitemOpen
  \bibfield  {author} {\bibinfo {author} {\bibfnamefont {Y.}~\bibnamefont
  {Lahini}}, \bibinfo {author} {\bibfnamefont {R.}~\bibnamefont {Pugatch}},
  \bibinfo {author} {\bibfnamefont {F.}~\bibnamefont {Pozzi}}, \bibinfo
  {author} {\bibfnamefont {M.}~\bibnamefont {Sorel}}, \bibinfo {author}
  {\bibfnamefont {R.}~\bibnamefont {Morandotti}}, \bibinfo {author}
  {\bibfnamefont {N.}~\bibnamefont {Davidson}},\ and\ \bibinfo {author}
  {\bibfnamefont {Y.}~\bibnamefont {Silberberg}},\ }\bibfield  {title}
  {\bibinfo {title} {Observation of a localization transition in quasiperiodic
  photonic lattices},\ }\href {https://doi.org/10.1103/PhysRevLett.103.013901}
  {\bibfield  {journal} {\bibinfo  {journal} {Phys. Rev. Lett.}\ }\textbf
  {\bibinfo {volume} {103}},\ \bibinfo {pages} {013901} (\bibinfo {year}
  {2009})}\BibitemShut {NoStop}%
\bibitem [{\citenamefont {Kraus}\ \emph {et~al.}(2012)\citenamefont {Kraus},
  \citenamefont {Lahini}, \citenamefont {Ringel}, \citenamefont {Verbin},\ and\
  \citenamefont {Zilberberg}}]{AAH7}%
  \BibitemOpen
  \bibfield  {author} {\bibinfo {author} {\bibfnamefont {Y.~E.}\ \bibnamefont
  {Kraus}}, \bibinfo {author} {\bibfnamefont {Y.}~\bibnamefont {Lahini}},
  \bibinfo {author} {\bibfnamefont {Z.}~\bibnamefont {Ringel}}, \bibinfo
  {author} {\bibfnamefont {M.}~\bibnamefont {Verbin}},\ and\ \bibinfo {author}
  {\bibfnamefont {O.}~\bibnamefont {Zilberberg}},\ }\bibfield  {title}
  {\bibinfo {title} {Topological states and adiabatic pumping in
  quasicrystals},\ }\href {https://doi.org/10.1103/PhysRevLett.109.106402}
  {\bibfield  {journal} {\bibinfo  {journal} {Phys. Rev. Lett.}\ }\textbf
  {\bibinfo {volume} {109}},\ \bibinfo {pages} {106402} (\bibinfo {year}
  {2012})}\BibitemShut {NoStop}%
\bibitem [{\citenamefont {Lang}\ \emph {et~al.}(2012)\citenamefont {Lang},
  \citenamefont {Cai},\ and\ \citenamefont {Chen}}]{AAH8}%
  \BibitemOpen
  \bibfield  {author} {\bibinfo {author} {\bibfnamefont {L.-J.}\ \bibnamefont
  {Lang}}, \bibinfo {author} {\bibfnamefont {X.}~\bibnamefont {Cai}},\ and\
  \bibinfo {author} {\bibfnamefont {S.}~\bibnamefont {Chen}},\ }\bibfield
  {title} {\bibinfo {title} {Edge states and topological phases in
  one-dimensional optical superlattices},\ }\href
  {https://doi.org/10.1103/PhysRevLett.108.220401} {\bibfield  {journal}
  {\bibinfo  {journal} {Phys. Rev. Lett.}\ }\textbf {\bibinfo {volume} {108}},\
  \bibinfo {pages} {220401} (\bibinfo {year} {2012})}\BibitemShut {NoStop}%
\bibitem [{\citenamefont {Ganeshan}\ \emph {et~al.}(2013)\citenamefont
  {Ganeshan}, \citenamefont {Sun},\ and\ \citenamefont {Das~Sarma}}]{AAH9}%
  \BibitemOpen
  \bibfield  {author} {\bibinfo {author} {\bibfnamefont {S.}~\bibnamefont
  {Ganeshan}}, \bibinfo {author} {\bibfnamefont {K.}~\bibnamefont {Sun}},\ and\
  \bibinfo {author} {\bibfnamefont {S.}~\bibnamefont {Das~Sarma}},\ }\bibfield
  {title} {\bibinfo {title} {Topological zero-energy modes in gapless
  commensurate aubry-andr\'e-harper models},\ }\href
  {https://doi.org/10.1103/PhysRevLett.110.180403} {\bibfield  {journal}
  {\bibinfo  {journal} {Phys. Rev. Lett.}\ }\textbf {\bibinfo {volume} {110}},\
  \bibinfo {pages} {180403} (\bibinfo {year} {2013})}\BibitemShut {NoStop}%
\bibitem [{\citenamefont {Verbin}\ \emph {et~al.}(2013)\citenamefont {Verbin},
  \citenamefont {Zilberberg}, \citenamefont {Kraus}, \citenamefont {Lahini},\
  and\ \citenamefont {Silberberg}}]{AAH10}%
  \BibitemOpen
  \bibfield  {author} {\bibinfo {author} {\bibfnamefont {M.}~\bibnamefont
  {Verbin}}, \bibinfo {author} {\bibfnamefont {O.}~\bibnamefont {Zilberberg}},
  \bibinfo {author} {\bibfnamefont {Y.~E.}\ \bibnamefont {Kraus}}, \bibinfo
  {author} {\bibfnamefont {Y.}~\bibnamefont {Lahini}},\ and\ \bibinfo {author}
  {\bibfnamefont {Y.}~\bibnamefont {Silberberg}},\ }\bibfield  {title}
  {\bibinfo {title} {Observation of topological phase transitions in photonic
  quasicrystals},\ }\href {https://doi.org/10.1103/PhysRevLett.110.076403}
  {\bibfield  {journal} {\bibinfo  {journal} {Phys. Rev. Lett.}\ }\textbf
  {\bibinfo {volume} {110}},\ \bibinfo {pages} {076403} (\bibinfo {year}
  {2013})}\BibitemShut {NoStop}%
\bibitem [{\citenamefont {Harper}(1955{\natexlab{b}})}]{AAHPT1}%
  \BibitemOpen
  \bibfield  {author} {\bibinfo {author} {\bibfnamefont {P.~G.}\ \bibnamefont
  {Harper}},\ }\bibfield  {title} {\bibinfo {title} {Single band motion of
  conduction electrons in a uniform magnetic field},\ }\href
  {https://doi.org/10.1088/0370-1298/68/10/304} {\bibfield  {journal} {\bibinfo
   {journal} {Proceedings of the Physical Society. Section A}\ }\textbf
  {\bibinfo {volume} {68}},\ \bibinfo {pages} {874} (\bibinfo {year}
  {1955}{\natexlab{b}})}\BibitemShut {NoStop}%
\bibitem [{\citenamefont {Sokoloff}(1985{\natexlab{a}})}]{AAHPT2}%
  \BibitemOpen
  \bibfield  {author} {\bibinfo {author} {\bibfnamefont {J.}~\bibnamefont
  {Sokoloff}},\ }\bibfield  {title} {\bibinfo {title} {Unusual band structure,
  wave functions and electrical conductance in crystals with incommensurate
  periodic potentials},\ }\href
  {https://doi.org/https://doi.org/10.1016/0370-1573(85)90088-2} {\bibfield
  {journal} {\bibinfo  {journal} {Physics Reports}\ }\textbf {\bibinfo {volume}
  {126}},\ \bibinfo {pages} {189} (\bibinfo {year}
  {1985}{\natexlab{a}})}\BibitemShut {NoStop}%
\bibitem [{\citenamefont {Longhi}(2021)}]{AAHPT3}%
  \BibitemOpen
  \bibfield  {author} {\bibinfo {author} {\bibfnamefont {S.}~\bibnamefont
  {Longhi}},\ }\bibfield  {title} {\bibinfo {title} {Phase transitions in a
  non-hermitian aubry-andr\'e-harper model},\ }\href
  {https://doi.org/10.1103/PhysRevB.103.054203} {\bibfield  {journal} {\bibinfo
   {journal} {Phys. Rev. B}\ }\textbf {\bibinfo {volume} {103}},\ \bibinfo
  {pages} {054203} (\bibinfo {year} {2021})}\BibitemShut {NoStop}%
\bibitem [{\citenamefont {Acharya}\ \emph {et~al.}(2022)\citenamefont
  {Acharya}, \citenamefont {Chakrabarty}, \citenamefont {Sahu},\ and\
  \citenamefont {Datta}}]{AAHPT4}%
  \BibitemOpen
  \bibfield  {author} {\bibinfo {author} {\bibfnamefont {A.~P.}\ \bibnamefont
  {Acharya}}, \bibinfo {author} {\bibfnamefont {A.}~\bibnamefont
  {Chakrabarty}}, \bibinfo {author} {\bibfnamefont {D.~K.}\ \bibnamefont
  {Sahu}},\ and\ \bibinfo {author} {\bibfnamefont {S.}~\bibnamefont {Datta}},\
  }\bibfield  {title} {\bibinfo {title} {Localization, $\mathcal{PT}$ symmetry
  breaking, and topological transitions in non-hermitian quasicrystals},\
  }\href {https://doi.org/10.1103/PhysRevB.105.014202} {\bibfield  {journal}
  {\bibinfo  {journal} {Phys. Rev. B}\ }\textbf {\bibinfo {volume} {105}},\
  \bibinfo {pages} {014202} (\bibinfo {year} {2022})}\BibitemShut {NoStop}%
\bibitem [{\citenamefont {Hiramoto}\ and\ \citenamefont
  {Abe}(1988)}]{dynamical_evolution1}%
  \BibitemOpen
  \bibfield  {author} {\bibinfo {author} {\bibfnamefont {H.}~\bibnamefont
  {Hiramoto}}\ and\ \bibinfo {author} {\bibfnamefont {S.}~\bibnamefont {Abe}},\
  }\bibfield  {title} {\bibinfo {title} {Dynamics of an electron in
  quasiperiodic systems. ii. harper's model},\ }\href
  {https://doi.org/10.1143/JPSJ.57.1365} {\bibfield  {journal} {\bibinfo
  {journal} {Journal of the Physical Society of Japan}\ }\textbf {\bibinfo
  {volume} {57}},\ \bibinfo {pages} {1365} (\bibinfo {year}
  {1988})}\BibitemShut {NoStop}%
\bibitem [{\citenamefont {Ketzmerick}\ \emph {et~al.}(1997)\citenamefont
  {Ketzmerick}, \citenamefont {Kruse}, \citenamefont {Kraut},\ and\
  \citenamefont {Geisel}}]{dynamical_evolution2}%
  \BibitemOpen
  \bibfield  {author} {\bibinfo {author} {\bibfnamefont {R.}~\bibnamefont
  {Ketzmerick}}, \bibinfo {author} {\bibfnamefont {K.}~\bibnamefont {Kruse}},
  \bibinfo {author} {\bibfnamefont {S.}~\bibnamefont {Kraut}},\ and\ \bibinfo
  {author} {\bibfnamefont {T.}~\bibnamefont {Geisel}},\ }\bibfield  {title}
  {\bibinfo {title} {What determines the spreading of a wave packet?},\ }\href
  {https://doi.org/10.1103/PhysRevLett.79.1959} {\bibfield  {journal} {\bibinfo
   {journal} {Phys. Rev. Lett.}\ }\textbf {\bibinfo {volume} {79}},\ \bibinfo
  {pages} {1959} (\bibinfo {year} {1997})}\BibitemShut {NoStop}%
\bibitem [{\citenamefont {Larcher}\ \emph {et~al.}(2009)\citenamefont
  {Larcher}, \citenamefont {Dalfovo},\ and\ \citenamefont
  {Modugno}}]{dynamical_evolution3}%
  \BibitemOpen
  \bibfield  {author} {\bibinfo {author} {\bibfnamefont {M.}~\bibnamefont
  {Larcher}}, \bibinfo {author} {\bibfnamefont {F.}~\bibnamefont {Dalfovo}},\
  and\ \bibinfo {author} {\bibfnamefont {M.}~\bibnamefont {Modugno}},\
  }\bibfield  {title} {\bibinfo {title} {Effects of interaction on the
  diffusion of atomic matter waves in one-dimensional quasiperiodic
  potentials},\ }\href {https://doi.org/10.1103/PhysRevA.80.053606} {\bibfield
  {journal} {\bibinfo  {journal} {Phys. Rev. A}\ }\textbf {\bibinfo {volume}
  {80}},\ \bibinfo {pages} {053606} (\bibinfo {year} {2009})}\BibitemShut
  {NoStop}%
\bibitem [{\citenamefont {Geisel}\ \emph {et~al.}(1991)\citenamefont {Geisel},
  \citenamefont {Ketzmerick},\ and\ \citenamefont
  {Petschel}}]{critical_behavior1}%
  \BibitemOpen
  \bibfield  {author} {\bibinfo {author} {\bibfnamefont {T.}~\bibnamefont
  {Geisel}}, \bibinfo {author} {\bibfnamefont {R.}~\bibnamefont {Ketzmerick}},\
  and\ \bibinfo {author} {\bibfnamefont {G.}~\bibnamefont {Petschel}},\
  }\bibfield  {title} {\bibinfo {title} {New class of level statistics in
  quantum systems with unbounded diffusion},\ }\href
  {https://doi.org/10.1103/PhysRevLett.66.1651} {\bibfield  {journal} {\bibinfo
   {journal} {Phys. Rev. Lett.}\ }\textbf {\bibinfo {volume} {66}},\ \bibinfo
  {pages} {1651} (\bibinfo {year} {1991})}\BibitemShut {NoStop}%
\bibitem [{\citenamefont {Machida}\ and\ \citenamefont
  {Fujita}(1986)}]{critical_behavior2}%
  \BibitemOpen
  \bibfield  {author} {\bibinfo {author} {\bibfnamefont {K.}~\bibnamefont
  {Machida}}\ and\ \bibinfo {author} {\bibfnamefont {M.}~\bibnamefont
  {Fujita}},\ }\bibfield  {title} {\bibinfo {title} {Quantum energy spectra and
  one-dimensional quasiperiodic systems},\ }\href
  {https://doi.org/10.1103/PhysRevB.34.7367} {\bibfield  {journal} {\bibinfo
  {journal} {Phys. Rev. B}\ }\textbf {\bibinfo {volume} {34}},\ \bibinfo
  {pages} {7367} (\bibinfo {year} {1986})}\BibitemShut {NoStop}%
\bibitem [{\citenamefont {Bertrand}\ and\ \citenamefont
  {Garc\'{\i}a-Garc\'{\i}a}(2016)}]{critical_behavior3}%
  \BibitemOpen
  \bibfield  {author} {\bibinfo {author} {\bibfnamefont {C.~L.}\ \bibnamefont
  {Bertrand}}\ and\ \bibinfo {author} {\bibfnamefont {A.~M.}\ \bibnamefont
  {Garc\'{\i}a-Garc\'{\i}a}},\ }\bibfield  {title} {\bibinfo {title} {Anomalous
  thouless energy and critical statistics on the metallic side of the many-body
  localization transition},\ }\href
  {https://doi.org/10.1103/PhysRevB.94.144201} {\bibfield  {journal} {\bibinfo
  {journal} {Phys. Rev. B}\ }\textbf {\bibinfo {volume} {94}},\ \bibinfo
  {pages} {144201} (\bibinfo {year} {2016})}\BibitemShut {NoStop}%
\bibitem [{\citenamefont {Lin}\ \emph {et~al.}(2023)\citenamefont {Lin},
  \citenamefont {Chen}, \citenamefont {Guo},\ and\ \citenamefont
  {Gong}}]{critical_behavior4}%
  \BibitemOpen
  \bibfield  {author} {\bibinfo {author} {\bibfnamefont {X.}~\bibnamefont
  {Lin}}, \bibinfo {author} {\bibfnamefont {X.}~\bibnamefont {Chen}}, \bibinfo
  {author} {\bibfnamefont {G.-C.}\ \bibnamefont {Guo}},\ and\ \bibinfo {author}
  {\bibfnamefont {M.}~\bibnamefont {Gong}},\ }\href@noop {} {\bibinfo {title}
  {The general approach to the critical phase with coupled quasiperiodic
  chains}} (\bibinfo {year} {2023}),\ \Eprint
  {https://arxiv.org/abs/2209.03060} {arXiv:2209.03060 [quant-ph]} \BibitemShut
  {NoStop}%
\bibitem [{\citenamefont {Halsey}\ \emph {et~al.}(1986)\citenamefont {Halsey},
  \citenamefont {Jensen}, \citenamefont {Kadanoff}, \citenamefont {Procaccia},\
  and\ \citenamefont {Shraiman}}]{multifractal_nature1}%
  \BibitemOpen
  \bibfield  {author} {\bibinfo {author} {\bibfnamefont {T.~C.}\ \bibnamefont
  {Halsey}}, \bibinfo {author} {\bibfnamefont {M.~H.}\ \bibnamefont {Jensen}},
  \bibinfo {author} {\bibfnamefont {L.~P.}\ \bibnamefont {Kadanoff}}, \bibinfo
  {author} {\bibfnamefont {I.}~\bibnamefont {Procaccia}},\ and\ \bibinfo
  {author} {\bibfnamefont {B.~I.}\ \bibnamefont {Shraiman}},\ }\bibfield
  {title} {\bibinfo {title} {Fractal measures and their singularities: The
  characterization of strange sets},\ }\href
  {https://doi.org/10.1103/PhysRevA.33.1141} {\bibfield  {journal} {\bibinfo
  {journal} {Phys. Rev. A}\ }\textbf {\bibinfo {volume} {33}},\ \bibinfo
  {pages} {1141} (\bibinfo {year} {1986})}\BibitemShut {NoStop}%
\bibitem [{\citenamefont {Mirlin}\ \emph {et~al.}(2006)\citenamefont {Mirlin},
  \citenamefont {Fyodorov}, \citenamefont {Mildenberger},\ and\ \citenamefont
  {Evers}}]{multifractal_nature2}%
  \BibitemOpen
  \bibfield  {author} {\bibinfo {author} {\bibfnamefont {A.~D.}\ \bibnamefont
  {Mirlin}}, \bibinfo {author} {\bibfnamefont {Y.~V.}\ \bibnamefont
  {Fyodorov}}, \bibinfo {author} {\bibfnamefont {A.}~\bibnamefont
  {Mildenberger}},\ and\ \bibinfo {author} {\bibfnamefont {F.}~\bibnamefont
  {Evers}},\ }\bibfield  {title} {\bibinfo {title} {Exact relations between
  multifractal exponents at the anderson transition},\ }\href
  {https://doi.org/10.1103/PhysRevLett.97.046803} {\bibfield  {journal}
  {\bibinfo  {journal} {Phys. Rev. Lett.}\ }\textbf {\bibinfo {volume} {97}},\
  \bibinfo {pages} {046803} (\bibinfo {year} {2006})}\BibitemShut {NoStop}%
\bibitem [{\citenamefont {Dubertrand}\ \emph {et~al.}(2014)\citenamefont
  {Dubertrand}, \citenamefont {Garc\'{\i}a-Mata}, \citenamefont {Georgeot},
  \citenamefont {Giraud}, \citenamefont {Lemari\'e},\ and\ \citenamefont
  {Martin}}]{multifractal_nature3}%
  \BibitemOpen
  \bibfield  {author} {\bibinfo {author} {\bibfnamefont {R.}~\bibnamefont
  {Dubertrand}}, \bibinfo {author} {\bibfnamefont {I.}~\bibnamefont
  {Garc\'{\i}a-Mata}}, \bibinfo {author} {\bibfnamefont {B.}~\bibnamefont
  {Georgeot}}, \bibinfo {author} {\bibfnamefont {O.}~\bibnamefont {Giraud}},
  \bibinfo {author} {\bibfnamefont {G.}~\bibnamefont {Lemari\'e}},\ and\
  \bibinfo {author} {\bibfnamefont {J.}~\bibnamefont {Martin}},\ }\bibfield
  {title} {\bibinfo {title} {Two scenarios for quantum multifractality
  breakdown},\ }\href {https://doi.org/10.1103/PhysRevLett.112.234101}
  {\bibfield  {journal} {\bibinfo  {journal} {Phys. Rev. Lett.}\ }\textbf
  {\bibinfo {volume} {112}},\ \bibinfo {pages} {234101} (\bibinfo {year}
  {2014})}\BibitemShut {NoStop}%
\bibitem [{\citenamefont {Wang}\ \emph {et~al.}(2016)\citenamefont {Wang},
  \citenamefont {Liu}, \citenamefont {Xianlong},\ and\ \citenamefont
  {Hu}}]{multifractal_nature4}%
  \BibitemOpen
  \bibfield  {author} {\bibinfo {author} {\bibfnamefont {J.}~\bibnamefont
  {Wang}}, \bibinfo {author} {\bibfnamefont {X.-J.}\ \bibnamefont {Liu}},
  \bibinfo {author} {\bibfnamefont {G.}~\bibnamefont {Xianlong}},\ and\
  \bibinfo {author} {\bibfnamefont {H.}~\bibnamefont {Hu}},\ }\bibfield
  {title} {\bibinfo {title} {Phase diagram of a non-abelian
  aubry-andr\'e-harper model with $p$-wave superfluidity},\ }\href
  {https://doi.org/10.1103/PhysRevB.93.104504} {\bibfield  {journal} {\bibinfo
  {journal} {Phys. Rev. B}\ }\textbf {\bibinfo {volume} {93}},\ \bibinfo
  {pages} {104504} (\bibinfo {year} {2016})}\BibitemShut {NoStop}%
\bibitem [{\citenamefont {Han}\ \emph {et~al.}(1994)\citenamefont {Han},
  \citenamefont {Thouless}, \citenamefont {Hiramoto},\ and\ \citenamefont
  {Kohmoto}}]{critically_localized_states1}%
  \BibitemOpen
  \bibfield  {author} {\bibinfo {author} {\bibfnamefont {J.~H.}\ \bibnamefont
  {Han}}, \bibinfo {author} {\bibfnamefont {D.~J.}\ \bibnamefont {Thouless}},
  \bibinfo {author} {\bibfnamefont {H.}~\bibnamefont {Hiramoto}},\ and\
  \bibinfo {author} {\bibfnamefont {M.}~\bibnamefont {Kohmoto}},\ }\bibfield
  {title} {\bibinfo {title} {Critical and bicritical properties of harper's
  equation with next-nearest-neighbor coupling},\ }\href
  {https://doi.org/10.1103/PhysRevB.50.11365} {\bibfield  {journal} {\bibinfo
  {journal} {Phys. Rev. B}\ }\textbf {\bibinfo {volume} {50}},\ \bibinfo
  {pages} {11365} (\bibinfo {year} {1994})}\BibitemShut {NoStop}%
\bibitem [{\citenamefont {Wang}\ \emph
  {et~al.}(2021{\natexlab{b}})\citenamefont {Wang}, \citenamefont {Cheng},
  \citenamefont {Liu},\ and\ \citenamefont
  {Yu}}]{critically_localized_states2}%
  \BibitemOpen
  \bibfield  {author} {\bibinfo {author} {\bibfnamefont {Y.}~\bibnamefont
  {Wang}}, \bibinfo {author} {\bibfnamefont {C.}~\bibnamefont {Cheng}},
  \bibinfo {author} {\bibfnamefont {X.-J.}\ \bibnamefont {Liu}},\ and\ \bibinfo
  {author} {\bibfnamefont {D.}~\bibnamefont {Yu}},\ }\bibfield  {title}
  {\bibinfo {title} {Many-body critical phase: Extended and nonthermal},\
  }\href {https://doi.org/10.1103/PhysRevLett.126.080602} {\bibfield  {journal}
  {\bibinfo  {journal} {Phys. Rev. Lett.}\ }\textbf {\bibinfo {volume} {126}},\
  \bibinfo {pages} {080602} (\bibinfo {year} {2021}{\natexlab{b}})}\BibitemShut
  {NoStop}%
\bibitem [{\citenamefont {Xiao}\ \emph
  {et~al.}(2021{\natexlab{b}})\citenamefont {Xiao}, \citenamefont {Xie},
  \citenamefont {Dong}, \citenamefont {Chen}, \citenamefont {Yi},\ and\
  \citenamefont {Yan}}]{critically_localized_states3}%
  \BibitemOpen
  \bibfield  {author} {\bibinfo {author} {\bibfnamefont {T.}~\bibnamefont
  {Xiao}}, \bibinfo {author} {\bibfnamefont {D.}~\bibnamefont {Xie}}, \bibinfo
  {author} {\bibfnamefont {Z.}~\bibnamefont {Dong}}, \bibinfo {author}
  {\bibfnamefont {T.}~\bibnamefont {Chen}}, \bibinfo {author} {\bibfnamefont
  {W.}~\bibnamefont {Yi}},\ and\ \bibinfo {author} {\bibfnamefont
  {B.}~\bibnamefont {Yan}},\ }\bibfield  {title} {\bibinfo {title} {Observation
  of topological phase with critical localization in a quasi-periodic
  lattice},\ }\href
  {https://doi.org/https://doi.org/10.1016/j.scib.2021.07.025} {\bibfield
  {journal} {\bibinfo  {journal} {Science Bulletin}\ }\textbf {\bibinfo
  {volume} {66}},\ \bibinfo {pages} {2175} (\bibinfo {year}
  {2021}{\natexlab{b}})}\BibitemShut {NoStop}%
\bibitem [{\citenamefont {Abanin}\ \emph {et~al.}(2019)\citenamefont {Abanin},
  \citenamefont {Altman}, \citenamefont {Bloch},\ and\ \citenamefont
  {Serbyn}}]{many-body_localization1}%
  \BibitemOpen
  \bibfield  {author} {\bibinfo {author} {\bibfnamefont {D.~A.}\ \bibnamefont
  {Abanin}}, \bibinfo {author} {\bibfnamefont {E.}~\bibnamefont {Altman}},
  \bibinfo {author} {\bibfnamefont {I.}~\bibnamefont {Bloch}},\ and\ \bibinfo
  {author} {\bibfnamefont {M.}~\bibnamefont {Serbyn}},\ }\bibfield  {title}
  {\bibinfo {title} {Colloquium: Many-body localization, thermalization, and
  entanglement},\ }\href {https://doi.org/10.1103/RevModPhys.91.021001}
  {\bibfield  {journal} {\bibinfo  {journal} {Rev. Mod. Phys.}\ }\textbf
  {\bibinfo {volume} {91}},\ \bibinfo {pages} {021001} (\bibinfo {year}
  {2019})}\BibitemShut {NoStop}%
\bibitem [{\citenamefont {Schreiber}\ \emph {et~al.}(2015)\citenamefont
  {Schreiber}, \citenamefont {Hodgman}, \citenamefont {Bordia}, \citenamefont
  {Lüschen}, \citenamefont {Fischer}, \citenamefont {Vosk}, \citenamefont
  {Altman}, \citenamefont {Schneider},\ and\ \citenamefont
  {Bloch}}]{many-body_localization2}%
  \BibitemOpen
  \bibfield  {author} {\bibinfo {author} {\bibfnamefont {M.}~\bibnamefont
  {Schreiber}}, \bibinfo {author} {\bibfnamefont {S.~S.}\ \bibnamefont
  {Hodgman}}, \bibinfo {author} {\bibfnamefont {P.}~\bibnamefont {Bordia}},
  \bibinfo {author} {\bibfnamefont {H.~P.}\ \bibnamefont {Lüschen}}, \bibinfo
  {author} {\bibfnamefont {M.~H.}\ \bibnamefont {Fischer}}, \bibinfo {author}
  {\bibfnamefont {R.}~\bibnamefont {Vosk}}, \bibinfo {author} {\bibfnamefont
  {E.}~\bibnamefont {Altman}}, \bibinfo {author} {\bibfnamefont
  {U.}~\bibnamefont {Schneider}},\ and\ \bibinfo {author} {\bibfnamefont
  {I.}~\bibnamefont {Bloch}},\ }\bibfield  {title} {\bibinfo {title}
  {Observation of many-body localization of interacting fermions in a
  quasirandom optical lattice},\ }\href
  {https://doi.org/10.1126/science.aaa7432} {\bibfield  {journal} {\bibinfo
  {journal} {Science}\ }\textbf {\bibinfo {volume} {349}},\ \bibinfo {pages}
  {842} (\bibinfo {year} {2015})}\BibitemShut {NoStop}%
\bibitem [{\citenamefont {Rispoli}\ \emph {et~al.}(2019)\citenamefont
  {Rispoli}, \citenamefont {Lukin}, \citenamefont {Schittko}, \citenamefont
  {Kim}, \citenamefont {Tai}, \citenamefont {L{\'e}onard},\ and\ \citenamefont
  {Greiner}}]{many-body_localization3}%
  \BibitemOpen
  \bibfield  {author} {\bibinfo {author} {\bibfnamefont {M.}~\bibnamefont
  {Rispoli}}, \bibinfo {author} {\bibfnamefont {A.}~\bibnamefont {Lukin}},
  \bibinfo {author} {\bibfnamefont {R.}~\bibnamefont {Schittko}}, \bibinfo
  {author} {\bibfnamefont {S.}~\bibnamefont {Kim}}, \bibinfo {author}
  {\bibfnamefont {M.~E.}\ \bibnamefont {Tai}}, \bibinfo {author} {\bibfnamefont
  {J.}~\bibnamefont {L{\'e}onard}},\ and\ \bibinfo {author} {\bibfnamefont
  {M.}~\bibnamefont {Greiner}},\ }\bibfield  {title} {\bibinfo {title} {Quantum
  critical behaviour at the many-body localization transition},\ }\href
  {https://doi.org/10.1038/s41586-019-1527-2} {\bibfield  {journal} {\bibinfo
  {journal} {Nature}\ }\textbf {\bibinfo {volume} {573}},\ \bibinfo {pages}
  {385} (\bibinfo {year} {2019})}\BibitemShut {NoStop}%
\bibitem [{\citenamefont {Jiang}\ \emph {et~al.}(2019)\citenamefont {Jiang},
  \citenamefont {Lang}, \citenamefont {Yang}, \citenamefont {Zhu},\ and\
  \citenamefont {Chen}}]{multicritical1}%
  \BibitemOpen
  \bibfield  {author} {\bibinfo {author} {\bibfnamefont {H.}~\bibnamefont
  {Jiang}}, \bibinfo {author} {\bibfnamefont {L.-J.}\ \bibnamefont {Lang}},
  \bibinfo {author} {\bibfnamefont {C.}~\bibnamefont {Yang}}, \bibinfo {author}
  {\bibfnamefont {S.-L.}\ \bibnamefont {Zhu}},\ and\ \bibinfo {author}
  {\bibfnamefont {S.}~\bibnamefont {Chen}},\ }\bibfield  {title} {\bibinfo
  {title} {Interplay of non-hermitian skin effects and anderson localization in
  nonreciprocal quasiperiodic lattices},\ }\href
  {https://doi.org/10.1103/PhysRevB.100.054301} {\bibfield  {journal} {\bibinfo
   {journal} {Phys. Rev. B}\ }\textbf {\bibinfo {volume} {100}},\ \bibinfo
  {pages} {054301} (\bibinfo {year} {2019})}\BibitemShut {NoStop}%
\bibitem [{\citenamefont {Longhi}(2019{\natexlab{b}})}]{multicritical2}%
  \BibitemOpen
  \bibfield  {author} {\bibinfo {author} {\bibfnamefont {S.}~\bibnamefont
  {Longhi}},\ }\bibfield  {title} {\bibinfo {title} {Topological phase
  transition in non-hermitian quasicrystals},\ }\href
  {https://doi.org/10.1103/PhysRevLett.122.237601} {\bibfield  {journal}
  {\bibinfo  {journal} {Phys. Rev. Lett.}\ }\textbf {\bibinfo {volume} {122}},\
  \bibinfo {pages} {237601} (\bibinfo {year} {2019}{\natexlab{b}})}\BibitemShut
  {NoStop}%
\bibitem [{\citenamefont {Liu}\ \emph {et~al.}(2021)\citenamefont {Liu},
  \citenamefont {Cheng}, \citenamefont {Guo},\ and\ \citenamefont
  {Xianlong}}]{multicritical3}%
  \BibitemOpen
  \bibfield  {author} {\bibinfo {author} {\bibfnamefont {T.}~\bibnamefont
  {Liu}}, \bibinfo {author} {\bibfnamefont {S.}~\bibnamefont {Cheng}}, \bibinfo
  {author} {\bibfnamefont {H.}~\bibnamefont {Guo}},\ and\ \bibinfo {author}
  {\bibfnamefont {G.}~\bibnamefont {Xianlong}},\ }\bibfield  {title} {\bibinfo
  {title} {Fate of majorana zero modes, exact location of critical states, and
  unconventional real-complex transition in non-hermitian quasiperiodic
  lattices},\ }\href {https://doi.org/10.1103/PhysRevB.103.104203} {\bibfield
  {journal} {\bibinfo  {journal} {Phys. Rev. B}\ }\textbf {\bibinfo {volume}
  {103}},\ \bibinfo {pages} {104203} (\bibinfo {year} {2021})}\BibitemShut
  {NoStop}%
\bibitem [{\citenamefont {Gavriliuk}\ \emph {et~al.}(2023)\citenamefont
  {Gavriliuk}, \citenamefont {Struzhkin}, \citenamefont {Ivanova},
  \citenamefont {Prakapenka}, \citenamefont {Mironovich}, \citenamefont
  {Aksenov}, \citenamefont {Troyan},\ and\ \citenamefont
  {Morgenroth}}]{first-order_pt1}%
  \BibitemOpen
  \bibfield  {author} {\bibinfo {author} {\bibfnamefont {A.~G.}\ \bibnamefont
  {Gavriliuk}}, \bibinfo {author} {\bibfnamefont {V.~V.}\ \bibnamefont
  {Struzhkin}}, \bibinfo {author} {\bibfnamefont {A.~G.}\ \bibnamefont
  {Ivanova}}, \bibinfo {author} {\bibfnamefont {V.~B.}\ \bibnamefont
  {Prakapenka}}, \bibinfo {author} {\bibfnamefont {A.~A.}\ \bibnamefont
  {Mironovich}}, \bibinfo {author} {\bibfnamefont {S.~N.}\ \bibnamefont
  {Aksenov}}, \bibinfo {author} {\bibfnamefont {I.~A.}\ \bibnamefont
  {Troyan}},\ and\ \bibinfo {author} {\bibfnamefont {W.}~\bibnamefont
  {Morgenroth}},\ }\bibfield  {title} {\bibinfo {title} {The first-order
  structural transition in nio at high pressure},\ }\href
  {https://doi.org/10.1038/s42005-022-01098-5} {\bibfield  {journal} {\bibinfo
  {journal} {Communications Physics}\ }\textbf {\bibinfo {volume} {6}},\
  \bibinfo {pages} {23} (\bibinfo {year} {2023})}\BibitemShut {NoStop}%
\bibitem [{\citenamefont {Liu}\ \emph {et~al.}(2023)\citenamefont {Liu},
  \citenamefont {Lu}, \citenamefont {Chu}, \citenamefont {Yang}, \citenamefont
  {Yuan}, \citenamefont {Wu}, \citenamefont {Ji}, \citenamefont {Tian},
  \citenamefont {Watanabe}, \citenamefont {Taniguchi}, \citenamefont {Du},
  \citenamefont {Shi}, \citenamefont {Liu}, \citenamefont {Shen}, \citenamefont
  {Lu}, \citenamefont {Yang},\ and\ \citenamefont {Zhang}}]{first-order_pt2}%
  \BibitemOpen
  \bibfield  {author} {\bibinfo {author} {\bibfnamefont {L.}~\bibnamefont
  {Liu}}, \bibinfo {author} {\bibfnamefont {X.}~\bibnamefont {Lu}}, \bibinfo
  {author} {\bibfnamefont {Y.}~\bibnamefont {Chu}}, \bibinfo {author}
  {\bibfnamefont {G.}~\bibnamefont {Yang}}, \bibinfo {author} {\bibfnamefont
  {Y.}~\bibnamefont {Yuan}}, \bibinfo {author} {\bibfnamefont {F.}~\bibnamefont
  {Wu}}, \bibinfo {author} {\bibfnamefont {Y.}~\bibnamefont {Ji}}, \bibinfo
  {author} {\bibfnamefont {J.}~\bibnamefont {Tian}}, \bibinfo {author}
  {\bibfnamefont {K.}~\bibnamefont {Watanabe}}, \bibinfo {author}
  {\bibfnamefont {T.}~\bibnamefont {Taniguchi}}, \bibinfo {author}
  {\bibfnamefont {L.}~\bibnamefont {Du}}, \bibinfo {author} {\bibfnamefont
  {D.}~\bibnamefont {Shi}}, \bibinfo {author} {\bibfnamefont {J.}~\bibnamefont
  {Liu}}, \bibinfo {author} {\bibfnamefont {J.}~\bibnamefont {Shen}}, \bibinfo
  {author} {\bibfnamefont {L.}~\bibnamefont {Lu}}, \bibinfo {author}
  {\bibfnamefont {W.}~\bibnamefont {Yang}},\ and\ \bibinfo {author}
  {\bibfnamefont {G.}~\bibnamefont {Zhang}},\ }\bibfield  {title} {\bibinfo
  {title} {Observation of first-order quantum phase transitions and
  ferromagnetism in twisted double bilayer graphene},\ }\href
  {https://doi.org/10.1103/PhysRevX.13.031015} {\bibfield  {journal} {\bibinfo
  {journal} {Phys. Rev. X}\ }\textbf {\bibinfo {volume} {13}},\ \bibinfo
  {pages} {031015} (\bibinfo {year} {2023})}\BibitemShut {NoStop}%
\bibitem [{\citenamefont {Dwivedi}\ and\ \citenamefont
  {Chua}(2016)}]{transfer_matrix1}%
  \BibitemOpen
  \bibfield  {author} {\bibinfo {author} {\bibfnamefont {V.}~\bibnamefont
  {Dwivedi}}\ and\ \bibinfo {author} {\bibfnamefont {V.}~\bibnamefont {Chua}},\
  }\bibfield  {title} {\bibinfo {title} {Of bulk and boundaries: Generalized
  transfer matrices for tight-binding models},\ }\href
  {https://doi.org/10.1103/PhysRevB.93.134304} {\bibfield  {journal} {\bibinfo
  {journal} {Phys. Rev. B}\ }\textbf {\bibinfo {volume} {93}},\ \bibinfo
  {pages} {134304} (\bibinfo {year} {2016})}\BibitemShut {NoStop}%
\bibitem [{\citenamefont {Cai}\ \emph {et~al.}(2019)\citenamefont {Cai},
  \citenamefont {Liu}, \citenamefont {Wu}, \citenamefont {He}, \citenamefont
  {Zhu}, \citenamefont {Zhang},\ and\ \citenamefont {Wang}}]{Haldane1}%
  \BibitemOpen
  \bibfield  {author} {\bibinfo {author} {\bibfnamefont {H.}~\bibnamefont
  {Cai}}, \bibinfo {author} {\bibfnamefont {J.}~\bibnamefont {Liu}}, \bibinfo
  {author} {\bibfnamefont {J.}~\bibnamefont {Wu}}, \bibinfo {author}
  {\bibfnamefont {Y.}~\bibnamefont {He}}, \bibinfo {author} {\bibfnamefont
  {S.-Y.}\ \bibnamefont {Zhu}}, \bibinfo {author} {\bibfnamefont {J.-X.}\
  \bibnamefont {Zhang}},\ and\ \bibinfo {author} {\bibfnamefont {D.-W.}\
  \bibnamefont {Wang}},\ }\bibfield  {title} {\bibinfo {title} {Experimental
  observation of momentum-space chiral edge currents in room-temperature
  atoms},\ }\href {https://doi.org/10.1103/PhysRevLett.122.023601} {\bibfield
  {journal} {\bibinfo  {journal} {Phys. Rev. Lett.}\ }\textbf {\bibinfo
  {volume} {122}},\ \bibinfo {pages} {023601} (\bibinfo {year}
  {2019})}\BibitemShut {NoStop}%
\bibitem [{\citenamefont {Sokoloff}(1985{\natexlab{b}})}]{Haldane2}%
  \BibitemOpen
  \bibfield  {author} {\bibinfo {author} {\bibfnamefont {J.}~\bibnamefont
  {Sokoloff}},\ }\bibfield  {title} {\bibinfo {title} {Unusual band structure,
  wave functions and electrical conductance in crystals with incommensurate
  periodic potentials},\ }\href
  {https://doi.org/https://doi.org/10.1016/0370-1573(85)90088-2} {\bibfield
  {journal} {\bibinfo  {journal} {Physics Reports}\ }\textbf {\bibinfo {volume}
  {126}},\ \bibinfo {pages} {189} (\bibinfo {year}
  {1985}{\natexlab{b}})}\BibitemShut {NoStop}%
\bibitem [{\citenamefont {Sachdev}(1999)}]{quantum_phase_transition}%
  \BibitemOpen
  \bibfield  {author} {\bibinfo {author} {\bibfnamefont {S.}~\bibnamefont
  {Sachdev}},\ }\bibfield  {title} {\bibinfo {title} {Quantum phase
  transitions},\ }\href {https://doi.org/10.1088/2058-7058/12/4/23} {\bibfield
  {journal} {\bibinfo  {journal} {Physics World}\ }\textbf {\bibinfo {volume}
  {12}},\ \bibinfo {pages} {33} (\bibinfo {year} {1999})}\BibitemShut {NoStop}%
\bibitem [{\citenamefont {Longhi}(2019{\natexlab{c}})}]{AAH1}%
  \BibitemOpen
  \bibfield  {author} {\bibinfo {author} {\bibfnamefont {S.}~\bibnamefont
  {Longhi}},\ }\bibfield  {title} {\bibinfo {title} {Topological phase
  transition in non-hermitian quasicrystals},\ }\href
  {https://doi.org/10.1103/PhysRevLett.122.237601} {\bibfield  {journal}
  {\bibinfo  {journal} {Phys. Rev. Lett.}\ }\textbf {\bibinfo {volume} {122}},\
  \bibinfo {pages} {237601} (\bibinfo {year} {2019}{\natexlab{c}})}\BibitemShut
  {NoStop}%
\bibitem [{\citenamefont {Ashida}\ \emph
  {et~al.}(2020{\natexlab{b}})\citenamefont {Ashida}, \citenamefont {Gong},\
  and\ \citenamefont {Ueda}}]{point_gap}%
  \BibitemOpen
  \bibfield  {author} {\bibinfo {author} {\bibfnamefont {Y.}~\bibnamefont
  {Ashida}}, \bibinfo {author} {\bibfnamefont {Z.}~\bibnamefont {Gong}},\ and\
  \bibinfo {author} {\bibfnamefont {M.}~\bibnamefont {Ueda}},\ }\bibfield
  {title} {\bibinfo {title} {Non-hermitian physics},\ }\href
  {https://doi.org/10.1080/00018732.2021.1876991} {\bibfield  {journal}
  {\bibinfo  {journal} {Advances in Physics}\ }\textbf {\bibinfo {volume}
  {69}},\ \bibinfo {pages} {249} (\bibinfo {year}
  {2020}{\natexlab{b}})}\BibitemShut {NoStop}%
\end{thebibliography}%

\end{document}